\documentclass[prb,twocolumn,noshowpacs,amsmath]{revtex4}
\usepackage{bm}
\usepackage{epsfig}
\begin{document}

\title{Self-consistent theory of current and voltage noise in multimode 
ballistic conductors}

\author{O. M. Bulashenko}
 \email{oleg@ffn.ub.es}
\author{J. M. Rub\'{\i}}
\affiliation{Departament de F\'{\i}sica Fonamental,
Universitat de Barcelona, Diagonal 647, E-08028 Barcelona, Spain}

\date{\today}

\begin{abstract}
Electron transport in a self-consistent potential along a ballistic 
two-terminal conductor has been investigated. 
We have derived general formulas which describe the nonlinear current-voltage 
characteristics, differential conductance, and low-frequency current and 
voltage noise assuming an arbitrary distribution function and correlation
properties of injected electrons.
The analytical results have been obtained for a wide range of biases:
from equilibrium to high values beyond the linear-response regime.
The particular case of a three-dimensional Fermi-Dirac injection has been 
analyzed. We show that the Coulomb correlations are manifested 
in the negative excess voltage noise, i.e., the voltage fluctuations
under high-field transport conditions can be less than in equilibrium.
\end{abstract}


\maketitle

\section{Introduction}

Recently, measurements of nonequilibrium noise have emerged as a fundamental 
tool to obtain information on the transport properties and interactions among 
carriers in mesoscopic systems. \cite{dejong97,landauer98,blanter00}
Shot-noise suppression in ballistic conductors caused by Fermi correlations 
has been studied extensively both theoretically
\cite{lesovik89,buttiker9092,land-martin9192}
and experimentally. \cite{li90,reznikov95,kumar96,brom99}
Within the scattering approach, it is usually assumed that the ballistic
(phase-coherent) conductor is attached to reservoirs (terminals or leads)
with different chemical potentials.
In this approach, the mean current in a two-terminal conductor is given by
\begin{equation}  \label{mcur1}
I = \frac{q}{2\pi\hbar} \sum_n \int d\varepsilon\ {\cal T}_n(\varepsilon) 
[f_L(\varepsilon)-f_R(\varepsilon)],
\end{equation}
where $q$ is the electron charge, $f_{L,R}(\varepsilon)$ the energy 
distribution functions at the left (L) and right (R) reservoirs, and 
${\cal T}_n$ the transmission probabilities associated with 
$n$ transverse quantum modes (channels). \cite{blanter00}
The corresponding current-noise power at zero-frequency has been obtained 
in the form \cite{lesovik89,buttiker9092} (also see 
Refs.~\onlinecite{blanter00,land-martin9192}, and \onlinecite{kogan}).
\begin{multline} \label{Stwo}
S_I = \frac{q^2}{\pi\hbar} \sum_n \int d\varepsilon\ \{ {\cal T}_n(\varepsilon)
[ f_L (1 - f_L) + f_R (1 - f_R) ] \\
+ {\cal T}_n(\varepsilon) [1 - {\cal T}_n(\varepsilon)] (f_L - f_R)^2 \}.
\end{multline}
In Eq.~(\ref{Stwo}) the noise is a combination of the thermal
emission noise of the reservoirs and of the partition noise appeared
due to the current partitioning between the incoming and outgoing states 
[scattering on tunneling barrier(s), elastic scatterer(s), point contact(s), 
etc.]. Although in some limits the well-known noise terms can be identified
(associated with thermal noise or shot noise), they cannot be separated 
in general.
Out of equilibrium, the noise can manifest itself in a different way,
depending on the conditions.
At low biases $U$, Eq.~(\ref{Stwo}) gives the {\em partition
shot noise}---the excess noise linear in current (bias),
which does not vanish at zero temperature. 
It is suppressed below the Poisson $2qI$ value
approximately by the factor \cite{blanter00} 
$\sum_n {\cal T}_n(1-{\cal T}_n)/\sum_n {\cal T}_n$.
This type of excess noise appears whenever there is a partitioning of current
(${\cal T}_n\neq 0;1$).
It vanishes for fully ballistic systems for which
there is no partitioning (${\cal T}_n$=1)
(see experimental evidence \cite{li90,reznikov95,kumar96,brom99}).
In the absence of partitioning, the excess noise is in general no longer 
linear in the current. 
The inherent randomness in the emission of carriers from the reservoirs
is at the origin of this type of nonequilibrium noise.
Presumably, it is more pronounced for sufficiently high biases when 
$f_R\ll f_L$ and the transmission dominates in only one direction.
In this case, noise formula (\ref{Stwo}) is simplified to 
\cite{buttiker9092}
\begin{equation}  \label{Stwo1}
S_I=\frac{q^2}{\pi\hbar}\sum_{n^*} \int f_L(1 - f_L)d\varepsilon,
\end{equation}
where the summation is taken for open channels. \cite{remark1}
For low electron densities the occupation numbers are small, $f_L\ll 1$, 
and  Eq.~(\ref{Stwo1}) leads to the Schottky formula 
$S_I=(q^2/\pi\hbar)\sum_{n^*} \int f_L d\varepsilon =2qI_{em}$, where $I_{em}$
is the emission current from the reservoir (vacuum-tube-like shot noise).
In this low-density limit, shot noise is Poissonian since 
the transmission of carriers is uncorrelated.
The factor $(1-f_L)$ in Eq.~(\ref{Stwo1}) introduces the Fermi correlations 
among carriers when the occupation numbers are not small in respect to 1.
This leads to the suppressed value of the shot noise.
Note the difference between the partition shot noise mentioned earlier,
and the {\em emission shot noise} given by Eq.~(\ref{Stwo1}).
The former persists at zero temperature, since it reflects the granularity
in charge transmission manifested by partitioning, while the latter vanishes 
at $T\to 0$, because its origin is the thermal fluctuations of the occupation 
numbers in the reservoirs.
Indeed, at equilibrium the sum of two opposite shot-noise terms 
[Eq.~(\ref{Stwo1})] gives the Nyquist formula. \cite{landauer89}

It should be stressed that both Eqs.~(\ref{Stwo}) and (\ref{Stwo1}) 
are not complete, since they ignore Coulomb interactions.
The electrons are charged entities and, while moving along the conductor, they
affect the electric potential giving rise to inhomogeneity. \cite{sablikov98}
The self-consistent coupling between the nonhomogeneous electron density and
potential landscape is very important to adequately describe the transport 
and noise under nonlinear far-from-equilibrium conditions. 
\cite{sablikov00,buttiker02}
An interesting question is how the self-consistency may affect the current 
and noise formulas (\ref{mcur1})--(\ref{Stwo1})?
First, the transmission probabilities ${\cal T}_n$ for both current and noise
become functionals of the {\em time-averaged} self-consistent potential 
profile $\bar{\varphi}$. \cite{wei99}
Second, in the current-noise formulas (\ref{Stwo}) and (\ref{Stwo1}), 
which reflect only the {\em injected} current fluctuations $\delta I_{inj}$,
the additional terms should appear caused by the current fluctuations 
$\delta I_{ind}$ {\em induced} by the {\em fluctuations} of the potential 
$\delta\varphi$.
Finding the fluctuations $\delta\varphi$ is a complicated problem
in general, since they are self-consistently linked to the fluctuations 
of the occupation numbers along the conductor. 
(The latter may be expressed through the fluctuations of the occupation 
numbers at the terminals, since the system is ballistic \cite{prb00a}). 
As a result of this self-consistent coupling, the long-range Coulomb 
correlations appear, which may result in the noise suppression.
\cite{land-martin9192,landauer9396,buttiker96}
It is believed, however, that such Coulomb correlations need
to be taken into account for the description of systems in time-dependent
external fields, or finite-frequency fluctuation spectra in stationary
fields, while the zero-frequency fluctuations in stationary 
fields are not affected by them. \cite{blanter00}
We show that this is not always true. In the example we address in this paper, 
the additional terms induced by the self-consistent field are of the order of 
the fluctuations injected from the leads and cannot be neglected even 
in the zero-frequency limit at time-independent biases. 
Moreover, {\em they can almost completely compensate the injected fluctuations 
up to an arbitrarily small value}. At the same time, the gauge invariance 
required for the charge conservation is fulfilled.
We also found that this Coulomb suppression of noise is manifested 
in the negative excess voltage noise.
Current or voltage fluctuations in equilibrium, described by 
the fluctuation-dissipation theorem, usually increase 
when an external electric field is applied.
We have an interesting example when
an interacting (via Coulomb forces) electron system is less noisy
at far-from-equilibrium conditions than in equilibrium.
For noninteracting systems, such examples have been given by Lesovik and 
Loosen. \cite{lesovik93}

To support our statement we present a theory of current and voltage 
fluctuations in a ballistic two-terminal conductor in a self-consistent field. 
(Fig.~1).
The calculation of the self-consistent fluctuating field is, in general, 
a multidimensional problem which includes the electrostatic environment. 
For simplicity, we consider rather thick samples that allow us
to use one-dimensional plane geometry for electrostatics.
On the other hand, for wide conductors the number of transversal modes is 
large and the semiclassical treatment is sufficient.
By assuming that there is no current partitioning 
(${\cal T}_n$=1 for all the transmitting modes),
we focus mainly on the nonequilibrium noise caused by thermal emission
from the reservoirs under the action of the long-range Coulomb correlations 
inside the ballistic region, rather than on the partition shot noise.
It should be noted that the previous theoretical studies have been devoted to 
ballistic conductors with a small number of quantum modes (quantum point 
contacts) with the Fermi suppression of the partition shot noise (Coulomb 
correlations have been ignored). 
\cite{lesovik89,buttiker9092,land-martin9192,scherbakov98}

The main results of the present investigation are as follows:
We have obtained complete analytical expressions for the steady-state
spatial profiles (carrier density, self-consistent field), mean current, and 
differential conductance, as well as the current and voltage noise powers
in ballistic multimode conductors.
The analytical results have been obtained for a wide range of biases:
from equilibrium to high values beyond the linear-response regime
under the self-consistent-field conditions.
We assume in our derivations arbitrary distribution functions and 
correlation properties of injected electrons in order to generalize the model 
to the practically important cases of nanoscale devices with nonequilibrium 
electron injection, like in a resonant-tunneling-diode emitter, superlattice 
emitter, hot-electron emitter, etc.\
(see, e.g., Refs.~\onlinecite{levi85,heiblum85,kast01,strahberger01}).
The particular case of a three-dimensional (3D) Fermi-Dirac injection 
has also been addressed. 
The obtained results clearly demonstrate that both the current and voltage 
noise can be substantially reduced owing to the long-range Coulomb 
interactions.
This result is very encouraging from the point of view of applications.

The paper is organized as follows.
In Sec.~II we introduce the basic equations describing 
the space-charge-limited (SCL) ballistic transport.
In Sec.~III the self-consistent steady-state spatial profiles for the electron
density, electric field and potential are found for an arbitrary injection
distribution. The mean current and conductance are obtained in Sec.~IV.
Section V describes a general formula which relates both current and
voltage fluctuations with the fluctuations of the occupation numbers 
in the leads. The current noise power, suppressed by interactions, 
is compared in Sec.~VI with the case when interactions are disregarded. 
The Coulomb and Fermi noise-suppression factors are discussed in Secs.~VII 
and VIII, respectively, whereas the noise temperature is given in Sec.~IX. 
The voltage noise power under a fixed-current conditions is derived in Sec.~X. 
The implementation of the results for a GaAs ballistic conductor is presented
in Sec.~XI. Finally, Sec.~XII summarizes the main contributions of the paper,
whereas in the Appendixes we present mathematical details concerning
some derivations.
\begin{figure}
\epsfxsize=8.0cm
\epsfbox{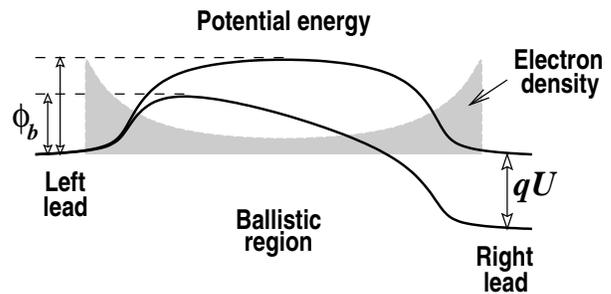}
\caption{Energy diagram determining the potential-barrier shape for a ballistic
two-terminal conductor at equilibrium and under applied bias $U$. 
Barrier height $\Phi_b$, as seen from the left lead, decreases with bias.
The filled area illustrates the nonhomogeneous electron-density distribution.}
\label{f1} \end{figure}

\section{Basic equations}

We consider a semiconductor ballistic sample attached to plane-parallel
leads (Fig.~\ref{f1}). 
In a semiclassical framework, the electron occupation numbers 
$\tilde{f}(x,{\bf k},t)$ inside the ballistic conductor are determined by 
the electron flows from the left and right leads.
The distribution of carriers is nonhomogeneous along the conductor: their 
concentration is higher near the leads and lower in the middle of the sample.
The inhomogeneity of the space charge disturbs the electrostatic potential 
in such a way that the self-consistent built-in field determines the potential 
barrier, at which electrons are either reflected or transmitted depending on 
their energy (Fig.~\ref{f1}). 
We neglect tunneling and quantum reflection, i.e., 
the transmission probability is 1 if the electron energy
is higher than the barrier height, and it is 0 in the opposite case.
Out of equilibrium, the barrier height is different 
for the left and right lead electrons.
If for the left electrons the barrier height is $\Phi_b$, for the right
electrons the barrier height is $\Phi_b+qU$ (Fig.~\ref{f1}).
This leads to an asymmetry in their contribution to the current:
as the bias is increased, the barrier for the left electrons progressively
decreases and the current from the left lead enhances,
whereas the barrier for the right electrons increases and the current from the 
right lead decreases disappearing at all at high biases.

The occupation numbers are described by the Vlasov equation (collisionless 
Boltzmann equation with a self-consistent field) \cite{prb00a,prb00b}
\begin{eqnarray} \label{kin}
\left(\frac{\partial}{\partial t} + 
\frac{\hbar k_x}{m} \frac{\partial}{\partial x} \right.
+ \left. q \frac{d\tilde{\varphi}}{d x} \frac{\partial}{\hbar\partial k_x}
\right) \tilde{f}(x,{\bf k},t)= 0, 
\end{eqnarray}
where $m$ is the electron effective mass, ${\bf k}=(k_x,{\bf k}_{\perp})$,
and $\tilde{\varphi}(x,t)$ is the self-consistent electric potential 
determined by the Poisson equation 
\begin{eqnarray} \label{vlasns}
\frac{\partial^2\tilde{\varphi}}{\partial x^2} = \frac{q}{\kappa} \,
\int  \frac{d{\bf k}}{(2\pi)^d}\ \tilde{f}(x,{\bf k},t).
\label{pois}
\end{eqnarray}
Here $\kappa$ is the dielectric permittivity and $d$ is the dimension of 
a momentum space (the spin variable is neglected).
Since carriers move without collisions, the only source of noise
arises from the random injection of carriers from the leads.
Thus the boundary conditions at the left (L) and right (R) leads are:
\begin{eqnarray} \label{bcvlas}
\tilde{f}(0,{\bf k},t)|_{k_x>0} &=& f_L({\bf k}) + \delta f_L({\bf k},t), 
\nonumber\\
\tilde{f}(\ell,{\bf k},t)|_{k_x<0} &=& f_R({\bf k}) + \delta f_R({\bf k},t), \\
\tilde\varphi(\ell,t) &-& \tilde\varphi(0,t) =  \tilde{U}(t), \nonumber
\end{eqnarray}
where $\delta f_{L,R}$ are the stochastic forces inside the leads 
with zero average and given correlation properties, 
and $\tilde{U}$ is the applied bias between $x$=0 and $x$=$l$
(the potential drop inside the leads is neglected).
As a consequence of stochastic injection, the occupation numbers
$\tilde{f}(x,{\bf k},t)=f(x,{\bf k})+\delta f(x,{\bf k},t)$ 
and hence the potential
$\tilde{\varphi}(x,t)=\varphi(x)+\delta\varphi(x,t)$ fluctuate in time
around their time-averaged values.

The leads are assumed to be completely absorptive, 
and the transverse electron momenta are conserved.
Thus, one can make summing up over the transversal states 
(the summation can be replaced by integration due to the assumption of 
a large number of modes)
and introduce for each longitudinal energy $\varepsilon$ the (fluctuating) 
occupation factor at a cross section $x$:
\begin{equation}  \label{nmod}
\tilde{n}(x,\varepsilon,t) = \int_0^{\infty} 
\tilde{f}(x,\varepsilon,\varepsilon_{\perp},t)\ 
\nu_{\perp}\ d\varepsilon_{\perp},
\end{equation}
where $\varepsilon$=$\hbar^2 k_x^2/(2m)$, $\varepsilon_{\perp}$=
$\hbar^2 {\bf k}_{\perp}^2/(2m)$, and $\nu_{\perp}$ is the density of 
transverse modes ($\nu_{\perp}=m/2\pi\hbar^2$ for the 3D case).
The number of occupied transversal modes is $N_{\perp}=nA$,
where $A$ is the cross-sectional area.
In the semiclassical description applied here for a thick conductor, 
the number of the occupied transversal modes is assumed to be large, 
$N_{\perp}\gg 1$.

\begin{figure}[b]
\epsfxsize=8.0cm
\epsfbox{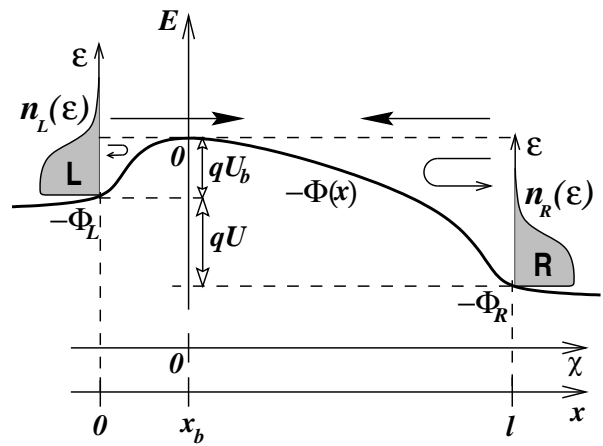}
\caption{
Potential-energy profile for a ballistic space-charge-limited conductor.
Electrons with energies $E>0$ pass over the barrier, while those
with $E<0$ are reflected back to the leads.
Shadowed regions illustrate the energy distributions of the occupation factors
at the leads.}
\label{f2} \end{figure}

\section{Self-consistent steady-state spatial profiles}
\label{sec-steady}

It is advantageous to introduce the mean total longitudinal energy $E$ as a 
sum of the kinetic energy $\varepsilon$ and the potential energy $[-\Phi(x)]$:
\begin{equation}
E=\varepsilon-\Phi(x).
\end{equation}
We shall count off the potential energy from the barrier top:
$\Phi(x)\equiv q\varphi(x)-q\varphi(x_b)$. Therefore, 
at the barrier position $x$=$x_b$, we obtain $E$=0 for electrons having the 
injection kinetic energy equal to the barrier height (for both leads).
The boundary values for the potential energy are expressed through the barrier 
height $qU_b$ and the applied bias $U$ as $\Phi_L\equiv\Phi(0)=qU_b$, and 
$\Phi_R\equiv\Phi(\ell)=\Phi_L+qU$ (Fig.~\ref{f2}).
The solution of Eq.~(\ref{kin}) for the stationary case
($\partial/\partial t$=0) gives, after integration over the transversal 
states, the electron density at any section of the conductor in terms of 
the potential $\Phi$, \cite{prb00a,prb00b}
\begin{widetext}
\begin{equation} \label{den}
N(\Phi)=
\int_0^{\infty} [n_L(E+\Phi_L)+n_R(E+\Phi_R)]
\nu(E+\Phi) dE 
 + 2\int_{-\Phi}^0 [\theta(-\chi)n_L(E+\Phi_L)
+\theta(\chi)n_R(E+\Phi_R)]
\nu(E+\Phi) dE, 
\end{equation}
\end{widetext}
where $n_{L,R}(E)=\int_0^{\infty} 
f_{L,R}(E,\varepsilon_{\perp}) \nu_{\perp} d\varepsilon_{\perp}$ are
the occupation factors at the leads, $\nu(E)=1/[2\pi\hbar v(E)]$ 
the density of states, and $v=\sqrt{2E/m}$ the velocity.
The first integral in Eq.~(\ref{den}) corresponds to the electrons 
transmitted over the barrier ($E>0$), while the second integral 
is referred to the reflected carriers ($-\Phi<E<0$). 
The latter term is doubled, since for each energy there are two states with 
opposite momentum: $k_x$ and $-k_x$.
Finally, $\theta(\chi)$ with $\chi=x-x_b$ is the Heaviside function that
distinguishes two classes of the reflected carriers: those at the left of 
the barrier ($\chi<0$) originated from the left lead, and those at the right 
($\chi>0$) coming form the right lead.

The electron density [Eq.~(\ref{den})] can now be substituted into the Poisson 
equation $d^2\Phi/dx^2=(q^2/\kappa)N(\Phi)$ to find the self-consistent 
potential $\Phi$. The first integration gives the electric-field distribution
\begin{eqnarray} \label{e}
{\cal E}(\Phi) = - \frac{1}{q}\frac{d\Phi}{dx} = -\text{sgn}(\chi)
\sqrt{\frac{2}{\kappa}}\sqrt{h(\Phi)},
\end{eqnarray}
where
\begin{widetext}
\begin{multline} \label{h}
h(\Phi) =
\int_0^{\Phi} N(\tilde{\Phi}) d\tilde{\Phi} 
=\frac{m}{2\pi\hbar}  \left\{ 
\int_0^{\infty} 
[n_L(E+\Phi_L)+n_R(E+\Phi_R)]
[v(E+\Phi)-v(E)] dE
\right. \\ 
\left. 
+  2\int_{-\Phi}^0 [\theta(-\chi)n_L(E+\Phi_L)
+\theta(\chi)n_R(E+\Phi_R)]
v(E+\Phi) dE \right\}.
\end{multline}
\end{widetext}
Integrating Eq.~(\ref{e}), one obtains the distribution of the potential
for both regions $\chi<0$ and $\chi>0$ in an implicit form
\begin{equation}\label{x} 
q\,\sqrt{\frac{2}{\kappa}} \, \chi = -\text{sgn}(\chi)
\int_0^{\Phi} 
\frac{d\tilde{\Phi}}{\sqrt{h(\tilde{\Phi})}}.
\end{equation}
Matching the two branches at $\chi=0$ yields
\begin{equation} \label{Vb}
q\,\ell \,\sqrt{\frac{2}{\kappa}} =
\int_0^{\Phi_L} \frac {d\Phi} {\sqrt{h_-(\Phi)}}
+ \int_0^{\Phi_R} \frac{d\Phi}{\sqrt{h_+(\Phi)}},
\end{equation}
with $h_-\equiv h(\chi<0)$ and $h_+\equiv h(\chi>0)$.
Equation (\ref{Vb}) relates three important parameters: the self-consistent 
barrier height $U_b$, the applied bias $U$, and the length of the conductor 
$\ell$. 
Given any two of them, the third one can be calculated \cite{remark2} from 
Eq.~(\ref{Vb}). In Ref.~\onlinecite{prb00a} a similar expression was obtained 
for the Maxwell-Boltzmann injection distribution. 
Here we have generalized it to the case of an arbitrary injection 
distribution profiles $n_L(E)$ and $n_R(E)$ at the leads.
It should be noted that in Eq.~(\ref{Vb}) the dependence on bias enters 
not only through the upper limits of the integrals, but also through 
the functions $h_{\pm}$. 

If we assume that electrons inside the leads obey the equilibrium Fermi-Dirac 
(FD) distributions (the usual assumption in other works) 
from Eq.~(\ref{nmod}) one obtains the occupation numbers
\begin{eqnarray} \label{nFD}
n_{L,R}(\varepsilon) = \frac{m k_B T}{2\pi\hbar^2}
\ln \{1+\exp[(\varepsilon_F -\varepsilon)/k_BT]\},
\end{eqnarray}
where $\varepsilon_F$ is the Fermi energy at the lead and
$T$ is the temperature.
Thus, for the FD case, the steady-state spatial profiles of the potential, 
electric field, and electron density are determined through 
Eqs.~(\ref{den})--(\ref{Vb}) by making use of the distributions
\begin{eqnarray} \label{nFD2}
n_L(E+\Phi_L) &=& 
\frac{\cal N}{\xi A}\, {\cal F}_0(\alpha-E/k_BT), \nonumber\\
n_R(E+\Phi_R) &=& 
\frac{\cal N}{\xi A}\, {\cal F}_0[\alpha-(E+qU)/k_BT],
\end{eqnarray}
where ${\cal N}=(k_F^2 A/4\pi)$ is the number of transversal modes
in the degenerate zero-temperature limit, 
$\xi=\varepsilon_F/k_BT$ is the reduced Fermi energy, 
$\alpha=(\varepsilon_F-\Phi_L)/k_BT$ is the parameter characterizing 
the position of the Fermi energy with respect to the potential barrier,
and ${\cal F}_k$ is the Fermi-Dirac integral of index $k$. \cite{prb00b}

\section{Mean current and conductance}

The mean ballistic current is found as an integral over the occupation numbers
for the transmitted ($E>0$) carriers from both leads:

\begin{eqnarray} \label{I}
I = \frac{qA}{2\pi\hbar} 
\int_0^{\infty} [n_L(E+\Phi_L) - n_R(E+\Phi_R)]\ dE.
\end{eqnarray}
It is convenient for future analysis to introduce the energy-resolved 
injection currents
\begin{eqnarray} \label{Ipart}
I_{L,R}(E) = \frac{qA}{2\pi\hbar} n_{L,R}(E).
\end{eqnarray}
By this definition, the mean current is 
\begin{eqnarray} \label{I1}
I = \int_0^{\infty} [I_L(E+\Phi_L) - I_R(E+\Phi_R)]\ dE. 
\end{eqnarray}
Having found the barrier height $\Phi_L$ from Eq.~(\ref{Vb}),
this equation determines the current-voltage characteristics.
Since we have not assumed that the bias must be small, 
this characteristics is nonlinear in bias in a general case.

For the FD case, Eq.~(\ref{I1}) reduces to
\begin{equation} \label{I3D}
I = \frac{q}{2\pi\hbar}\ {\cal N}\ \frac{k_BT}{\xi}\
[{\cal F}_1(\alpha) - {\cal F}_1(\alpha-V)],
\end{equation}
where we have denoted the dimensionless bias $V=qU/(k_BT)$.
It is seen that under the ballistic SCL conduction,
the current is determined by the relative positions of the Fermi energies
at the leads and the barrier top through the parameters $\alpha$ and $V$.
This is in contrast to the case of diffusive conductors,
in which the current is determined by the scattering strength.

The differential conductance $G=dI/dU$ is obtained from Eq.~(\ref{I}) as 
\begin{equation}\label{g} 
G =  \frac{q^2}{2\pi\hbar} A 
\left\{n_R(\Phi_R) - \frac{d U_b}{dU} \,[n_L(\Phi_L) - n_R(\Phi_R] \right\}.
\end{equation}
The derivative $dU_b/dU$ is calculated in the Appendix \ref{app-g}. 
With its help the formula for the conductance becomes
\begin{equation}\label{g1} 
G = \frac{q^2}{2\pi\hbar} A \left[ n_L(\Phi_L) \frac{\Delta_R}{\Delta}
+n_R(\Phi_R) \frac{\Delta_L}{\Delta} \right].
\end{equation}
It is seen that the conductance is a sum of two contributions corresponding
to the left and right leads. Each of them is a product of the conductance unit
$G_0=q^2/(2\pi\hbar)$, the number of the transversal modes 
for the injection energy corresponding to the barrier top, and some
Coulomb interaction factors determined through $\Delta_{L,R}$
given in Appendix \ref{app-g}. These factors depends on the whole electron
system and cannot be separated for the left and right lead electrons.

At small biases close to equilibrium, by assuming identical leads 
(e.g., FD distributions), we obtain $\Delta_{L,R}\approx\Delta/2$.
For this case the interaction factors vanish, and the conductance reduces to 
the value given by the multichannel Landauer formula
\begin{equation}\label{geq} 
G_{eq} \approx \frac{q^2}{2\pi\hbar} N_{\perp}(\Phi_L^0).
\end{equation}
where $N_{\perp}(\Phi_L^0)=A n_L(\Phi_L^0)$ is the number of open modes 
at the barrier energy. Under this small-bias condition, the current-voltage 
characteristics is linear: $I\approx G_{eq} U$.
Equation (\ref{g1}) may be viewed as the extension of the Landauer
formula for the conductance to far-from-equilibrium conditions 
for interacting electrons in a SCL ballistic conductor.

In the opposite limit of high biases $U_b\ll U<U_{cr}$, where $U_{cr}$ is the 
critical bias under which the barrier vanishes, 
the asymptotic formula for the current is \cite{prb00b}
\begin{equation} \label{Ias}
I_A \approx I_{\text{Child}}
\left[ 1 + \frac{3}{\sqrt{\Phi_R}} 
\frac{\int_0^{\infty} I_L(E+\Phi_L)\ \sqrt{E}\ dE}
{\int_0^{\infty} I_L(E+\Phi_L)\ dE} \right],
\end{equation}
where the leading-order term is the Child current
\begin{equation} \label{Ich}
I_{\text{Child}} = 
\frac{4}{9}\kappa A\sqrt{\frac{2}{m}}\frac{\Phi_R^{3/2}}{q\,l^2}.
\end{equation}
The main term $\propto U^{3/2}$ is independent of the injection, 
while the second-order term $\propto U$ contains information on 
the injection occupation numbers. Equation (\ref{Ias}) for the FD case 
was presented in Ref.~\onlinecite{prb01}.

The asymptotic behavior of the conductance at high biases is obtained from 
Eq.~(\ref{Ias}) 
\begin{equation}\label{gA} 
G_A = \frac{3}{2} \frac{qI_{\text{Child}}}{\Phi_R}
\left[ 1 + \frac{2}{\sqrt{\Phi_R}} 
\frac{\int_0^{\infty} I_L(E+\Phi_L)\ \sqrt{E}\ dE}
{\int_0^{\infty} I_L(E+\Phi_L)\ dE} \right],
\end{equation}
giving the leading-order term 
$G_A\approx(3/2)(I_{\text{Child}}/U) \sim\sqrt{U}$ for an arbitrary injection.
For the FD case, Eq.~(\ref{gA}) leads to
\begin{equation}\label{gAFD} 
G_A = \frac{3}{2} \, \frac{I_{\text{Child}}}{U}
\left[ 1 +  \sqrt{\frac{\pi k_BT}{qU}} 
\frac{{\cal F}_{3/2}(\alpha)}{{\cal F}_1(\alpha)} \right],
\end{equation}

\section{Self-consistent current and voltage fluctuations. General formulas}
\label{sec-gen}

According to the definition of the potential energy in Sec.~\ref{sec-steady}, 
the potential fluctuations at any point $x$ are given by 
$\delta\Phi_x = q\delta\varphi(x)-q\delta\varphi(x_b)$.
In the nonstationary frame fixed to the barrier top, the fluctuations 
at the barrier position are zero, $\delta\Phi_{x_b}=0$, whereas at the leads 
they are: $\delta\Phi_0\equiv\delta\Phi_L$ and 
$\delta\Phi_{\ell}\equiv\delta\Phi_R=\delta\Phi_L+q\delta U$, where 
$\delta U$ is the fluctuation of the applied bias.

The current fluctuation is obtained by integrating over the energy 
the fluctuation of the occupation factor $\delta n(E)$
found from a linearization of Eq.~(\ref{kin}) around 
the mean values. \cite{prb00b} One obtains
\begin{multline} \label{dI}
\delta I =
\int_0^{\infty} [\delta I_L(E+\Phi_L)-\delta I_R(E+\Phi_R)] \, d E \\ 
- I_L(\Phi_L) \, \delta\Phi_L + I_R(\Phi_R) \, \delta\Phi_R.
\end{multline}
where $\delta I_{L,R}(E)$ are the energy-resolved injection-current 
fluctuations from each lead.
In Eq.~(\ref{dI}), only the low-frequency current fluctuations are considered,
i.e., the frequencies are below the inverse electron transit time
between the leads and the displacement current is neglected.
The first integral term is standard, and corresponds to the injected current 
fluctuation $\delta I_{inj}$.
The last two terms are the fluctuations induced by the self-consistent 
potential fluctuations, that give rise to the long-range Coulomb correlations.
\cite{prb00b}
To find those terms, we need to obtain $\delta\Phi_L$ or, equivalently,
the self-consistent fluctuations of the barrier height in terms of 
the injected fluctuations $\delta I_{L,R}$ by solving the Poisson equation. 
This has been done in the Appendix \ref{app-pot}. The result is the relation
\begin{multline} \label{rel3}
\delta I - G \delta U =
\int_{-\Phi_L}^{\infty} \gamma_L(E) \, \delta I_L(E+\Phi_L) \, d E \\
+ \int_{-\Phi_R}^{\infty} \gamma_R(E) \, \delta I_R(E+\Phi_R) \, d E,
\end{multline}
where $G$ is the differential conductance [Eq.~(\ref{g1})]
and the functions $\gamma_{L,R}(E)$ are determined by
\begin{widetext}
\begin{equation}
\begin{aligned} \label{gamlr}
\gamma_L(E) &= \begin{cases} 
- 2C_{\Delta}  {\displaystyle
\int_{-E}^{\Phi_L}\frac{v(E+\Phi)}{h_-^{3/2}}\,d\Phi }, & -\Phi_L < E < 0 \\
1 - C_{\Delta} \left[ {\displaystyle \int_0^{\Phi_L}
\frac{v(E+\Phi) - v(E)}{h_-^{3/2}}\,d\Phi } +
{\displaystyle \int_0^{\Phi_R}\frac{v(E+\Phi) 
- v(E)}{h_+^{3/2}}\,d\Phi }\right],                     & 0 < E < \infty,
\end{cases}  \\ 
\gamma_R(E) &= \begin{cases} 
- 2C_{\Delta} {\displaystyle 
\int_{-E}^{\Phi_R}\frac{v(E+\Phi)}{h_+^{3/2}}\,d\Phi }, & -\Phi_R < E < 0  \\
-1 - C_{\Delta} \left[ {\displaystyle \int_0^{\Phi_L}
\frac{v(E+\Phi) - v(E)}{h_-^{3/2}}\,d\Phi } +
{\displaystyle \int_0^{\Phi_R}\frac{v(E+\Phi) 
- v(E)}{h_+^{3/2}}\,d\Phi }\right],                     & 0 < E < \infty. 
\end{cases}
\end{aligned}
\end{equation}
\end{widetext}
In Eq.~(\ref{gamlr}), one can distinguish the contributions from the
left-lead ($\gamma_L$) and right-lead ($\gamma_R$) electrons, as well as
from the reflected ($E<0$) and transmitted ($E>0$) carriers.
All the terms related to the barrier fluctuations are proportional to 
the constant
\begin{equation} \label{Cdelta}
C_{\Delta} = \frac{m}{2\pi\hbar\,\Delta} \, [n_L(\Phi_L)-n_R(\Phi_R)],
\end{equation}
where $\Delta$ is the constant previously used to determine the conductance 
$G$ and which has been derived in the Appendix \ref{app-g}. Equations 
(\ref{rel3})--(\ref{Cdelta}) are one of the main results of our theory.
They relate the self-consistent current and voltage fluctuations
with the noise source---spontaneous fluctuations of the occupation
numbers in the leads. The transfer functions $\gamma_{L,R}$, summarizing
the interaction effects, show the contribution of each energy to the
total fluctuations. In the absence of interactions, 
$\gamma_L(E)=\theta(E)$ and $\gamma_R(E)=-\theta(E)$, i.e., the fluctuations 
of all energies above the barrier top are equally transmitted. 
The role of the Coulomb interactions is to introduce an inhomogeneity
in the energy flux of fluctuations, by suppressing or enhancing 
occupation-number fluctuations at different energies.
Note that the Coulomb interactions are pronounced only in the presence 
of transport. In equilibrium, $C_{\Delta}=0$, and they are not effective.

In general, both terms $\gamma_L$ and $\gamma_R$ may contribute to the noise.
However, at high biases, $U_b \ll U < U_{cr}$, one can find that only
$\gamma_L(E>0)$ dominates, the asymptotic expression for which is
given by \cite{prb00b}
\begin{eqnarray} \label{gamma_asymp}
\gamma_L^A(E) = \frac{3}{\sqrt{\Phi_R}} 
\left( \sqrt{E} - v_{\Delta} \right) 
+ O\left(\frac{1}{\Phi_R}\right),
\end{eqnarray}
\begin{equation}
v_{\Delta} = \frac{1}{n_L(\Phi_L)} \int_0^{\infty} 
\left[- \frac{\partial n_L(E+\Phi_L)}{\partial E} \right] \sqrt{E}\ dE.
\end{equation}
We shall use these formulas later on to analyze the asymptotic limits for 
other important noise quantities.

To find the total fluctuations $\delta I$ or $\delta U$, one needs to define
the correlation properties of the fluctuations at the leads.
In general, one can write \cite{prb00b}
\begin{equation} \label{kuncor}
\langle\delta I_k(E)\delta I_k(E')\rangle = 
K_k(E)(\Delta f)\delta(E-E'),
\end{equation}
where, $k=L,R$ and $\Delta f$ is the frequency bandwidth 
(we assume the low-frequency limit).
For the case of the Poissonian injection from both leads
$K_{L,R}(E)\propto I_{L,R}(E)$. 
More generally, for the non-Poissonian injection, under the assumption that 
the leads are in local equilibrium, one can use the formula 
\begin{equation} \label{K}
K_{L,R}(E) = 2 k_B T\ G_0 A 
\left(-\frac{\partial n_{L,R}}{\partial E}\right),
\end{equation}
where $G_0$ is the unit of conductance. This formula follows from 
the Nyquist theorem (see Appendix \ref{app-N}).

By applying Eq.~(\ref{K}) for the FD case, we also obtain
\begin{equation} \label{KF3D}
K_{L,R}(E) = \frac{2 G_S}{\xi}\ \frac{1}{1+e^{\xi-(E/k_BT)}},
\end{equation}
where $G_S=G_0{\cal N}$ is the Sharvin conductance, and
$G_0=q^2/(2\pi\hbar)$ is the unit of conductance. 
For further noise analysis, we have to specify the conditions imposed on 
the external circuit. We shall consider two cases of interest:
(i) a voltage controlled circuit (zero external impedance) for which
$\delta U=0$ and one can find the spectral density of current fluctuations
$S_I$; and (ii) a current controlled circuit (infinite external impedance)
for which $\delta I=0$ and one can find the spectral density of voltage
fluctuations $S_V$ on the leads. 
In both cases, we will show that Coulomb interactions play a prominent role
in the noise suppression.


\section{Coulomb suppressed current noise}

Let us suppose that the potentials at the leads are held fixed and 
do not fluctuate. This corresponds to the case when currents are measured 
using a zero-impedance external circuit. Under the condition $\delta U=0$, 
Eq.~(\ref{rel3}) gives the current-noise spectral density
\begin{multline} \label{SI}
S_I = \int_{-\Phi_L}^{\infty} \gamma_L^2(E)\,
K_L(E+\Phi_L)\,dE \\
+ \int_{-\Phi_R}^{\infty} \gamma_R^2(E)\,
K_R(E+\Phi_R)\,dE.
\end{multline}
It is important to highlight that the obtained current-noise power 
[Eq.~(\ref{SI})],
that includes Coulomb interactions,  has been obtained for a wide range of 
biases, ranging from equilibrium to far-from-equilibrium conditions beyond 
the linear-response regime. 
Therefore, it describes both thermal and shot-noise limits.

One can verify that Eq.~(\ref{SI}) at the high-bias limit $U_b \ll U < U_{cr}$
reduces to
\begin{equation} \label{SIAgen}
S_I^A =  \int_0^{\infty} \gamma_L^2(E)\,
K_L(E+\Phi_L)\,dE.
\end{equation}
Taking into account Eqs.~(\ref{gamma_asymp}) and (\ref{K}), one obtains 
the asymptotic expression for the current-noise power 
\begin{gather} \label{SIA}
S_I^A \approx  \beta \,2qI\, \frac{k_BT}{qU}
= \frac{\beta}{3}\, 4k_BT\,G_A, \\
\intertext{where}
\beta = 9 \left(1 - \frac{\left[\int_0^{\infty} I_L(E+\Phi_L) 
\displaystyle{\frac{dE}{2\sqrt{E}}}\right]^2}
{ I_L(\Phi_L) \int_0^{\infty} I_L(E+\Phi_L)dE} \right).
\label{beta}
\end{gather}
The parameter $\beta$ is determined by the energy profile of the injected 
electrons $I_L(E)$. For the FD injection, Eq.~(\ref{beta})
leads to the formula derived earlier \cite{prb01}:
\begin{eqnarray} \label{betaFD}
\beta(\alpha) = 9 \left(1 - \frac{\pi}{4}
\frac{[{\cal F}_{1/2}(\alpha)]^2}{{\cal F}_0(\alpha){\cal F}_1(\alpha)}\right).
\end{eqnarray}
Equation (\ref{beta}) is more general and can be applied to an arbitrary 
injection distribution obeying the Nyquist relationship [Eq.~(\ref{K})]
for the correlation function. It is seen also from Eq.~(\ref{SIA}), that 
at high biases $S_I^A\sim \sqrt{U}$.

One can also find, for comparison, the current-noise power for the case of 
disregarded Coulomb correlations
\begin{equation*} \label{Siuncor0}
S_I^{\text{uncor}} = 
\int_0^{\infty} K_L(E+\Phi_L) \,dE + \int_0^{\infty} K_R(E+\Phi_R) \,dE,
\end{equation*}
which under the assumption of equilibrium conditions at the leads
[Eq.~(\ref{K})] results in
\begin{equation} \label{Siuncor}
S_I^{\text{uncor}} = 2qk_B T\ [I_L(\Phi_L)+I_R(\Phi_R)].
\end{equation}
For the sake of completeness, we present also expression for the FD case:
\begin{eqnarray} \label{SiuncorFD}
S_I^{\text{uncor}} = 2k_BT G_S\ \frac{1}{\xi}\
[{\cal F}_0(\alpha) + {\cal F}_0(\alpha-V)].
\end{eqnarray}
Note that Eq.~(\ref{SiuncorFD}) corresponds to Eq.~(\ref{Stwo1}) 
discussed in Sec.~I.
Indeed, if one applies Eq.~(\ref{Stwo1}) for two opposite flows of 
noninteracting FD electrons, summing up over the open channels,
one then gets Eq.~(\ref{SiuncorFD}).

\section{Coulomb noise-suppression factor}

To estimate the significance of Coulomb interactions, one can introduce 
the Coulomb noise-suppression factor \cite{prb01}
\begin{equation} \label{supC}
\Gamma_C =
\frac{S_I}{S_I^{\text{uncor}}},
\end{equation}
that extends over both thermal-noise and shot-noise limits. 
Strictly in equilibrium, $\Gamma_C=1$, as was pointed out 
in Sec.~\ref{sec-gen}. The effect of interactions is noticeable, however, 
already under small applied biases.
In Sec.~\ref{resultsA}, we will show that while $S_I^{\text{uncor}}$ 
increases with bias, the behavior of $S_I$ is just the opposite:
it decreases with bias starting from $U$=0 up to a certain bias where
it reaches the noise minimum, then $S_I$ increases but much slower
than $S_I^{\text{uncor}}$.

We remark the difference between the {\em noise}-suppres\-sion factor 
[Eq.~(\ref{supC})] and 
the {\em shot-noise}-suppression factor (also referred to as the Fano factor 
\cite{blanter00}) given by
\begin{equation} \label{fano}
F = \frac{S_I^{ex}}{2qI}.
\end{equation}
In the latter formula, the noise power $S_I^{ex}$ refers to 
the shot-noise power, i.e., the {\em excess} to the thermal-equilibrium-noise 
level: \cite{land-martin9192} $S_I^{ex}=S_I-4k_BT G_{eq}$. Moreover, 
Eq.~(\ref{fano}) is meaningful for systems in which $S_I^{ex}\propto I$ 
(for instance, in linear-response regime). In this case, it simply gives 
a measure of how much the noise power deviates from the ideal Poissonian $2qI$
value due to correlations among carriers.
For the nonlinear case, when $S_I^{ex}$ is not proportional to $I$, 
definition (\ref{fano}) is less useful, since the suppression factor 
depends on $I$.
It should be noted that Eqs.~(\ref{supC}) and (\ref{fano}) become identical
under the conditions: $qU\gg k_BT$ (for negligible thermal-noise contribution)
and $S_I^{\text{uncor}}=2qI$. The latter is valid, 
for instance, for the Maxwell-Boltzmann nondegenerate injection. \cite{prb00a}
If the injection is non-Poissonian, as in the case of FD injection, 
$S_I^{\text{uncor}}\neq 2qI$, and Eqs.~(\ref{supC}) and (\ref{fano}) differ.

\section{Fermi noise-suppression factor}
\label{secGF}

It is instructive to introduce the Poissonian noise power for the full range
of biases:
\begin{eqnarray} \label{SIP}
S_I^P &=& 2qI \coth\left(\frac{qU}{2k_BT}\right) \\
& \approx & \begin{cases}
4k_BT G_{eq}, & qU \ll k_BT \\
2qI, & qU \gg k_BT. \nonumber
\end{cases}
\end{eqnarray}
Based on this definition, one can introduce the Fermi noise-suppression factor
\begin{equation} \label{supF}
\Gamma_F = \frac{S_I^{\text{uncor}}}{S_I^P}.
\end{equation}
Thus the total noise-suppression factor is
\begin{equation} \label{sup}
\Gamma = \Gamma_F \Gamma_C = \frac{S_I}{S_I^P},
\end{equation}
Note that all definitions (\ref{supC}), (\ref{supF}), and (\ref{sup}) 
extend from thermal- to shot-noise limits.
They will be used in the analysis of the results in Sec.~\ref{results}.

For the FD two-lead injection, from Eqs.~(\ref{I3D}) and (\ref{SiuncorFD}) 
it follows that
\begin{equation} \label{supF-FD}
\Gamma_F = \frac{{\cal F}_0(\alpha) + {\cal F}_0(\alpha-V)}
{{\cal F}_1(\alpha) - {\cal F}_1(\alpha-V)} \tanh(V/2).
\end{equation}
One can verify that at the low-bias limit $V\to 0$, there is no suppression 
effect: $\Gamma_F\to 1$.
The finite bias introduces asymmetry in the contributions from electrons 
of different leads. The larger the bias, the smaller is the contribution 
from the biased lead, since the electrons have a higher potential barrier 
by an additional amount of $V$ to overcome. It is clear that starting from 
a certain bias, the contribution from only one injecting lead dominates.
The unidirectional charge flow occurs when 
\begin{equation} \label{condV}
V\agt\max\{5;\alpha+3\}.
\end{equation}
This condition comes from the consideration of two limits.
For a nondegenerate injection (the Fermi energy is below the barrier, 
$\alpha\alt -3$) the bias-to-temperature ratio should be large: 
$qU\agt 5k_BT$, whereas for a highly degenerate injection (the Fermi energy 
is above the barrier, $\alpha\agt 3$), the bias should be compared 
with the Fermi energy: $qU\agt \varepsilon_F - qU_b + 3k_BT$.
Thus, under condition (\ref{condV}), from Eq.~(\ref{supF-FD}) one obtains
the asymptotic formula \cite{prb01}
\begin{equation} \label{gaFas} 
\Gamma_F \approx \frac{{\cal F}_0(\alpha)}{{\cal F}_1(\alpha)}
\approx \begin{cases}
1, & \alpha\alt -3 \\
\displaystyle{\frac{2}{\alpha + 3\pi/\alpha^2},}
& \alpha\agt 3.
\end{cases}
\end{equation} 

It is also of interest to analyze the case opposite to condition 
(\ref{condV}), when the bias is not so high that both leads contribute
to the charge flow, namely, $V\alt\max\{5;\alpha+3\}$.
Then, for a nondegenerate limit $\Gamma_F\approx 1$.
For a highly degenerate limit, that happens when $V\alt\alpha-3$, 
one can use the approximations for the Fermi-Dirac integrals 
\cite{prb01} ${\cal F}_0(x)\approx x$ and 
${\cal F}_1(x)\approx x^2/2+\pi^2/6$, and 
Eq.~(\ref{supF-FD}) leads to a simple formula
\begin{equation} \label{supF-FDd}
\Gamma_F = \frac{2}{V} \tanh\left(\frac{V}{2}\right).
\end{equation}
Surprisingly, the dependence on $\alpha$ and hence on the barrier height,
ballistic length, and material parameters, canceled out from this equation.
$\Gamma_F$ depends only on one parameter---the bias-to-temperature ratio,
and at sufficiently high $V$ it decreases with bias as $\Gamma_F\approx 2/V$.
Note that this behavior occurs under the nonlinear bias regime
in the presence of a space charge. 
Indeed, for this case one finds the sublinear characteristics for the mean 
current $I=G_S U [1-q(U_b+U/2)/\varepsilon_F]$, whereas the current noise power
is given by $S_I^{\text{uncor}}=4k_BT G_S [1-q(U_b+U/2)/\varepsilon_F]$, 
with the identical factor in square brackets.
As a result, one obtains, $S_I^{\text{uncor}}=4k_BT\,I/U$,
as in the linear-response regime.
It should be remembered, however, that at higher biases $V\gg\alpha$, 
the $2/V$ dependence is changed to the $2/\alpha$ law.
The largest noise suppression by Fermi correlations \cite{prb01} is described 
by Eq.~(\ref{gaFas}) giving $\Gamma_F^{\text{min}}=2/\xi$, when the barrier
height is zero. We shall give some examples in Sec.~\ref{results}.

\section{Noise temperature}

It is interesting to see from Eq.~(\ref{SIA}), that at high biases and 
strong screening, despite the strong nonlinearity, the ratio between $S_I$ 
and the differential conductance $G$ tends to the constant value. 
It is instructive, therefore, to introduce the effective noise temperature 
$T_n$ through \cite{ziel86} $S_I = 4k_BT_n G$. 
Note that $k_B T_n$ has a meaning of the maximum noise power per unit 
bandwidth which can be delivered to an output matched circuit, 
thus it is a measurable quantity. \cite{shiktorov96}
The  asymptotic high-bias value is then obtained as
\begin{equation} \label{Tn}
\frac{T_n}{T} = \frac{1}{3}\beta(\alpha)
\approx \begin{cases}
3(1-\pi/4), & \alpha\alt -3 \\
1/3, & \alpha\gg 1.
\end{cases}
\end{equation}
It is seen that $T_n<T$ for any $\alpha$, indicating the noise suppression
effect [see the plot of $\beta(\alpha)$ in Ref.~\onlinecite{prb01}].
For a nondegenerate Maxwell-Boltzmann injection, the limiting value
$(T_n^{\text{ndeg}}/T)=3(1-\pi/4)\approx0.644$ is well known.
\cite{ziel86,apl99} For a highly degenerate FD injection, 
we have obtained from our theory $(T_n^{\text{deg}}/T)=1/3$.
The physical meaning of the latter result is that the noise power per unit 
bandwidth produced by the SCL ballistic conductor with degenerate FD electrons
is 1/3 of the thermal noise power produced by the heated resistance with 
the same value of the conductance $G$ (the same $I$-$V$ dependence),
independently of the material parameters.

\section{Coulomb suppressed voltage noise}
\label{secSU}

Alternatively, one could measure the voltages at the leads using an ideal 
infinite-impedance voltmeter. The infinite-impedance external circuit then 
forces the current to be zero at all times, $\delta I=0$.
Fluctuations in the current are counterbalanced by fluctuations of 
the chemical potentials in the electron reservoirs.
Under the condition $\delta I=0$, Eq.~(\ref{rel3}) gives the voltage-noise
spectral density which takes into account the Coulomb correlations:
\begin{multline} \label{SU}
S_U =
\frac{1}{G^2} \left[ \int_{-\Phi_L}^{\infty} \gamma_L^2(E)\,
K_L(E+\Phi_L)\,dE \right. \\
\left. +\int_{-\Phi_R}^{\infty} \gamma_R^2(E)\,
K_R(E+\Phi_R)\,dE \right].
\end{multline}
It is evident that the relation
\begin{equation} \label{SU1}
S_U = \frac{S_I}{G^2} 
\end{equation}
holds, in which $S_I$ is the current-noise spectral density [Eq.~(\ref{SI})]
measured under $\delta U=0$, and $G$ is the steady-state differential
conductance (\ref{g1}). 

The asymptotic behavior of $S_U$ at high biases can also be found
from Eqs.~(\ref{SIA}) and (\ref{SU1}). We obtain
\begin{equation} \label{SUas}
S_U^A = \frac{\beta}{3}\, 4k_BT\,\frac{1}{G_A},
\end{equation}
with $\beta$ given by Eq.~(\ref{beta}). It is seen that 
the voltage noise decreases with bias as $S_U^A\sim 1/\sqrt{U}$ at 
$U_b \ll U < U_{cr}$ (the general result independent of the injection
distribution).
Hence the Coulomb interactions result in the voltage-noise suppression.
This fact will be discussed in detail in Sec.~\ref{resultsB}.


\section{Results for Fermi-Dirac injection}
\label{results}

To illustrate our results, consider the GaAs ballistic $n$-$i$-$n$ diode 
at $T$=4 K. \cite{remark3}
For this temperature and the effective mass $m$=0.067$m_0$, 
the effective density of states is
$N_c\approx 6.7\times 10^{14}$ $\text{cm}^{-3}$.
Assuming the contact doping $1.6\times 10^{16}\,\text{cm}^{-3}$, the reduced
Fermi energy $\xi \approx 10$, and the contact electrons are
degenerate, that is necessary for studying the joint effect of both
Fermi and Coulomb correlations.
For this set of parameters, the Debye screening length associated with
the contact degenerate electron density is approximately
$L_D=\sqrt{\kappa k_BT/[q^2 N_c {\cal F}_{-1/2}(\xi)]}\approx 14$ nm.
The calculations have been carried out for the following ballistic lengths:
$\ell$=0.05, 0.1, and 0.5$\,\mu$m.

\begin{figure}[h]
\epsfxsize=8.0cm
\epsfbox{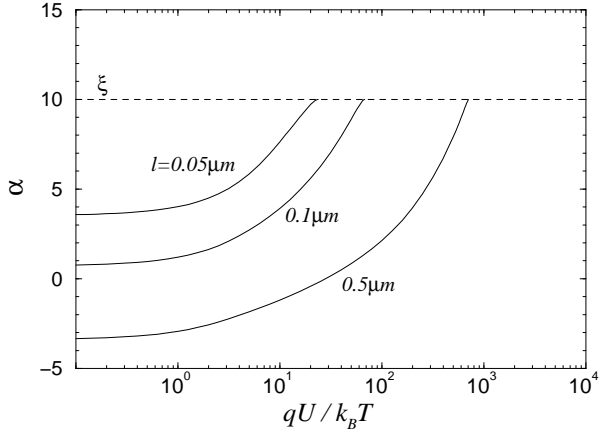}
\caption{Parameter $\alpha=(\varepsilon_F-\Phi_L)/k_BT$ 
characterizing the position of the Fermi energy 
$\varepsilon_F$ with respect to the potential barrier $\Phi_L$
for different lengths of the ballistic region $\ell$.
At $U\to U_{cr}$ the barrier vanishes, $\Phi_L$=0, and $\alpha$
attains its maximum value $\alpha$=$\xi$=10.}
\label{f3} \end{figure}

The degeneracy of the contact electrons does not guarantee the degeneracy of 
the injection, since the potential barrier determines the energy portion of 
electrons which may pass over the barrier and contribute to the injection 
current.
For each ballistic length $\ell$ and bias $U$, we have solved numerically 
Eq.~(\ref{Vb}) to find the self-consistent potential barrier height
$\Phi_L$ and the parameter $\alpha=(\varepsilon_F-\Phi_L)/k_BT$ 
characterizing the position of the Fermi energy with respect to the
potential barrier. The results are plotted in Fig.~\ref{f3}.

It is seen that in equilibrium, for the ballistic lengths 
$\ell$=0.05, 0.1, and 0.5$\,\mu$m, the self-consistent barrier height $\Phi_L$
is about 6 $k_BT$, 9 $k_BT$, and 13 $k_BT$, respectively.
This means that the injected electrons at $U$=0 are degenerate for 
$\ell$=0.05 and  0.1$\,\mu$m, and nondegenerate for $\ell$=0.5$\,\mu$m,
since for the latter case only the tail of the distribution function 
is injected ($\alpha<-3$). 
As $U$ is increased, $\Phi_L$ vanishes and the injection becomes degenerate 
for all three cases.
Finally, at $U\to U_{cr}$ the potential barrier vanishes, $\alpha$=$\xi$,
and the transport is no longer space-charge limited.
The values of $U_{cr}$ depend obviously on the ballistic length $\ell$ 
(see Fig.~\ref{f3}).
\begin{figure}[b]
\epsfxsize=8.0cm
\epsfbox{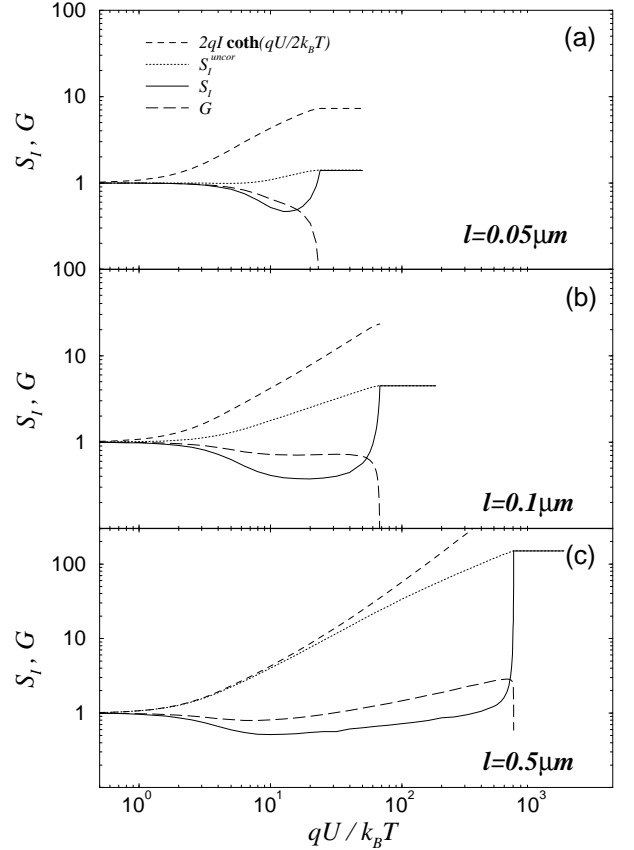}
\caption{Current-noise power $S_I$ and differential conductance $G$ 
vs bias $U$ for different ballistic lengths $\ell$.
For comparison the noise power for the Poissonian injection 
$2qI \coth(qU/2k_BT)$ and the Fermi-Dirac injection with disregarded 
Coulomb correlations $S_I^{\text{uncor}}$ are plotted.
The normalization constants are the corresponding equilibrium values
at $U$=0: $S_I^{eq}=4k_BT G_{eq}$ and $G_{eq}$.}
\label{f4} \end{figure}

\subsection{Fixed-bias conditions: $\bm{\delta V=0}$}
\label{resultsA}

In Fig.~\ref{f4} we show the results for the current-noise power $S_I$ 
[Eq.~(\ref{SI})] and differential conductance $G$ vs bias $U$ calculated 
for different ballistic lengths $\ell$. 
For comparison, the noise power for the Poissonian injection $S_I^P$ 
[Eq.~(\ref{SIP})] and Fermi-Dirac injection with disregarded Coulomb 
correlations $S_I^{\text{uncor}}$ [Eq.~(\ref{SiuncorFD})] have also been 
plotted.
In equilibrium, all the noise-power curves coalesce toward the Johnson-Nyquist 
noise $S_I^{eq}=4k_BT G_{eq}$ independently of the presence of Fermi or 
Coulomb correlations.
However, starting from small biases the difference becomes drastic.
While $S_I^{\text{uncor}}$ increases with bias,
the behavior of $S_I$ is just the opposite:
it decreases with bias starting from $U$=0 up to a certain bias where
it reaches the noise minimum, then $S_I$ increases, but much slower
than $S_I^{\text{uncor}}$.
Finally, at $U\to U_{cr}$ when the barrier vanishes, $S_I$ sharply
recovers $S_I^{\text{uncor}}$.
Note that in the absence of Coulomb interactions, $S_I^{\text{uncor}}$ follows
the Poissonian law $S_I^P=2qI \coth(qU/2k_BT)$ only for a nondegenerate 
injection, as in the case of $\ell$=0.5$\,\mu$m at $qU\lesssim 10 k_BT$ 
[Fig.~\ref{f4}(c)].
At higher biases, and for shorter ballistic lengths in all the range, the 
injection is non-Maxwellian and $S_I < S_I^P$  because of Fermi suppression 
[Figs.~\ref{f4}(a) and \ref{f4}(b)].
It should also be noted that $S_I < S_I^{eq}$ in a wide range of biases,
which means that the noise for interacting ballistic electrons in an external 
field is less than the equilibrium Johnson-Nyquist noise.
As will be shown later, the same is true for the voltage noise power.

\begin{figure}[b]
\epsfxsize=8.0cm
\epsfbox{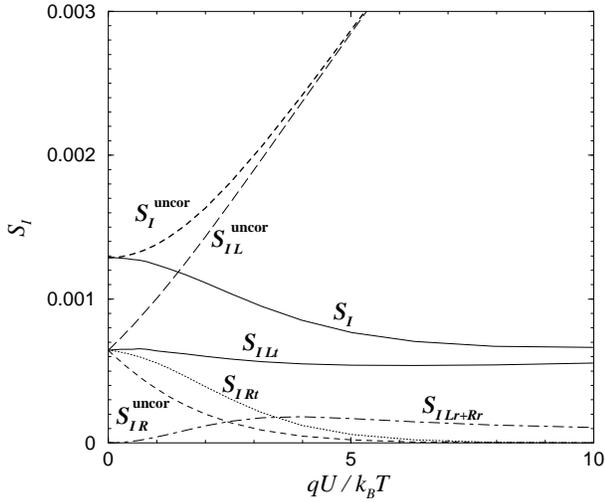}
\caption{Contributions to the current-noise power $S_I$ for the case of 
$\ell$=0.5$\mu$m:
from overbarrier electrons transmitted from the left ($S_{I\,Lt}$) and 
right ($S_{I\,Rt}$) leads, and those reflected by the barrier ($S_{I\,Lr+Rr}$).
For comparison, contributions to $S_I^{\text{uncor}}$ are shown as well: 
from left- ($S_{I\,L}^{\text{uncor}}$) and right-lead 
($S_{I\,R}^{\text{uncor}}$) electrons.}
\label{f5} \end{figure}

Figure \ref{f5} assists the understanding of the results by showing 
the contributions to $S_I$ from different electron groups.
The effect of Coulomb correlations is manifested quite differently for
the left-lead and right-lead electrons: while the left-lead noise is suppressed
($S_{I Lt}<S_{I L}^{\text{uncor}}$), the right-lead noise is enhanced
($S_{I Rt}>S_{I R}^{\text{uncor}}$). Since the role of the right-lead electrons
is diminished with bias, the overall effect of interaction results in
the total-noise suppression. There is also a non-negligible contribution 
($\sim 10-15\%$) to the noise from the reflected carriers at $qU\alt 10k_BT$. 
It appears for correlated electrons only.
\begin{figure}
\epsfxsize=8.0cm
\epsfbox{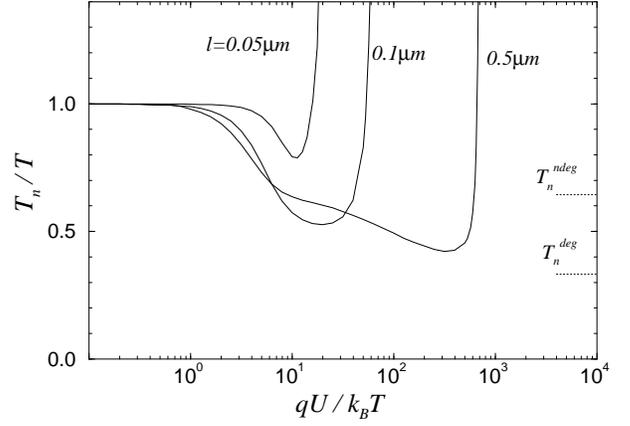}
\caption{Noise temperature $T_n$ vs bias $U$ for different ballistic lengths 
$\ell$. The asymptotic ($\ell\to\infty$, $U\to\infty$) limits for 
nondegenerate and highly degenerate electrons are shown: 
$T_n^{ndeg}=3(1-\pi/4)T$ and $T_n^{deg}=(1/3)T$, respectively.}
\label{f6} \end{figure}

Figure \ref{f6} shows the noise temperature $T_n$ versus bias $U$ calculated 
from $(T_n/T)=S_I/(4k_BTG)$ by using the data of Fig.~\ref{f4}. 
One can see that starting from $T_n$=$T$ at zero bias it drops at 
$qU\agt k_BT$ below the temperature $T$ of the injected electrons. 
It is interesting to note that for degenerate electrons ($\ell$=0.05$\,\mu$m),
this drop starts to appear at higher biases than for nondegenerate electrons 
($\ell$=0.5$\,\mu$m).
According to Eq.~(\ref{Tn}), the minimal asymptotic value of $T_n$ 
in the limit $\ell\to\infty$, $U\to\infty$ differs for nondegenerate and 
degenerate electrons (see indications in Fig.~\ref{f6}).
For our set of parameters, the injection is degenerate at the highest
biases. However, the limit $T_n^{deg}=(1/3)T$ is not achieved for those
ballistic lengths, since the samples are not sufficiently long.
Note that $T_n$ is a measurable quantity, and the observation of $T_n<T$
would indicate the significance of the Coulomb correlations effects 
which suppress the current noise.
At $U\sim U_{cr}$, $T_n$ sharply increases due to the current 
saturation ($G$=0), that may also be detected in the experiment.

\begin{figure}
\epsfxsize=8.5cm
\epsfbox{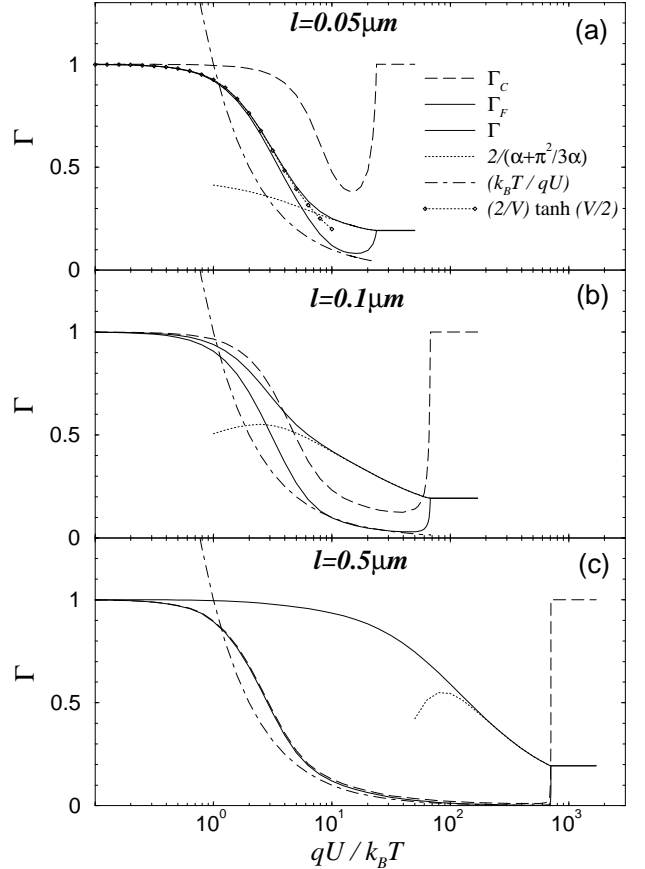}
\caption{Current-noise-suppression factors $\Gamma_C$ (Coulomb), 
$\Gamma_F$ (Fermi), and $\Gamma$=$\Gamma_F\Gamma_C$ (total) 
as functions of applied bias $U$ for different ballistic lengths $\ell$.
The analytical approximations $2/(\alpha+\pi^2/3\alpha)$ and 
$2/V\tanh(V/2)$ for $\Gamma_F$ and $k_BT/qU$ for $\Gamma$ are also shown.}
\label{f7} \end{figure}

The current-noise-suppression factors $\Gamma_C$ and $\Gamma_F$, and the total 
$\Gamma$ (their multiplication) correspondent to the noise-power curves of 
Fig.~\ref{f4} are plotted in Fig.~\ref{f7} as functions of bias.
The behavior of the Fermi suppression factor $\Gamma_F$ is in a close agreement
with the analytical formulas of Sec.~\ref{secGF}.
It varies from 1 at low biases, then decreases attaining with a nice
precision the asymptotic dependence $2/(\alpha+3\pi/\alpha^2)$ at high
biases [Eq.~(\ref{sup})]. 
For all three cases, the same minimal value $2/\xi\approx 0.2$ is reached 
at the highest biases, in agreement with the predictions. \cite{prb01}
We have also checked that for the degenerate injection from both leads, 
which is well realized for the length $\ell$=0.05$\,\mu$m, the analytical 
formula [Eq.~(\ref{supF-FDd})] very nicely describes the numerical results 
in a wide bias range: from $U$=0 up to $U\sim 5k_BT$ [Fig.~\ref{f7}(a)].
\noindent

The relative significance of two mechanisms on the noise suppression can be 
understood by comparing the curves for three different ballistic lengths 
$\ell$. For short samples ($\ell$=0.05$\,\mu$m), the Fermi suppression 
dominates at low biases $V\alt 3$, where $\Gamma_C\approx 1$ and 
$\Gamma\approx\Gamma_F$ [see Fig.~\ref{f7}(a)].
At higher biases, $3\alt V\alt 20$,
both Coulomb and Fermi mechanisms contribute to the suppression.
For $\ell$=0.1$\,\mu$m, the suppression factors $\Gamma_F$ and $\Gamma_C$
are comparable in all the bias range.
Finally, for the longer sample, $\ell$=0.5$\,\mu$m, the Coulomb noise 
suppression completely dominates: $\Gamma_C\ll\Gamma_F$.
This behavior can be explained by the fact that 
the Fermi shot-noise-suppression factor is limited below 
by the value $2/\xi$, i.e., by the properties of the injecting contact 
independently of the ballistic length. \cite{prb01}
In contrast, the Coulomb noise suppression may be enhanced arbitrarily strong 
by extending the length of the ballistic sample with a simultaneous increase
of bias (provided the transport remains ballistic).
Therefore, for any degree of electron degeneracy $\xi=\varepsilon_F/k_BT$, 
there exists the ballistic length starting from which the Coulomb interactions 
become to dominate in the noise suppression.

Another important difference between the two suppression mechanisms is that
the Fermi noise-suppression factor $\Gamma_F$ is a monotonically decreasing 
function of bias, while the Coulomb noise-suppression factor $\Gamma_C$ 
exhibits a minimum at a certain bias value, as seen in Fig.~\ref{f7}. 
After the minimum, the curve of $\Gamma_C$ increases to 1 due to 
the disappearance of the potential barrier at $U=U_{cr}$. 

The total noise-suppression factor $\Gamma$ approaches at high biases
the asymptotic curve $k_BT/qU$, once the injection, because of barrier 
lowering, becomes fully degenerate, in agreement with the prediction. 
\cite{prb01} The longer the sample, the wider is the bias range 
in which this asymptotic law is fulfilled independently of the material 
parameters (Fig.~\ref{f7}). 
It is also important that the suppression may be several orders of magnitude 
stronger than the shot-noise suppression due to elastic partitioning.  
\cite{blanter00}
\begin{figure}
\epsfxsize=8.0cm
\epsfbox{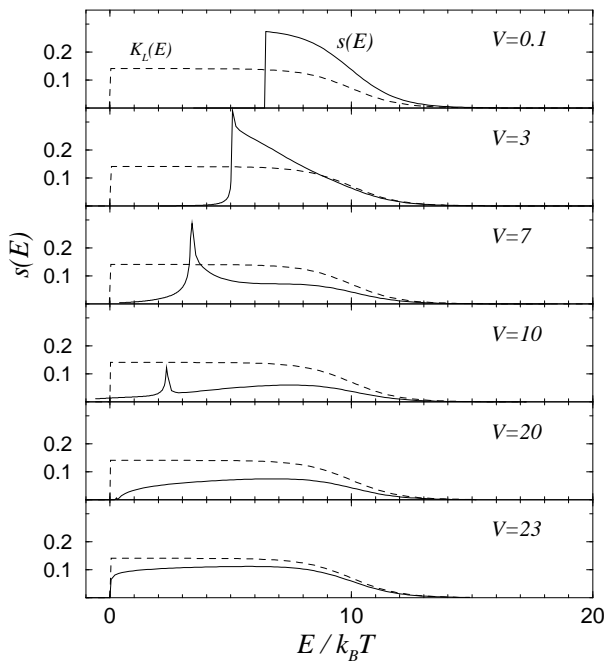}
\caption{Energy distributions of the current-noise power $s(E)$ 
under various biases $V$ for a ballistic conductor with $\ell$=0.05$\mu$m 
(solid lines). The zero energy corresponds to the conduction-band edge
at the left (unbiased) lead. The sharp peak at low energies corresponds to 
the position of the space-charge barrier. 
The dashed line shows the energy profile $K_L(E)$ for the injection noise
at the left lead ($\varepsilon_F=10k_BT$). The profile for the right lead
$K_R(E)$ is the same, but is shifted by $-V$ in energy.
All the curves are normalized by $4qI_L(\Phi_L^0)$ related to the noise
level at $V$=0.}
\label{f8} \end{figure}
\begin{figure}
\epsfxsize=8.0cm
\epsfbox{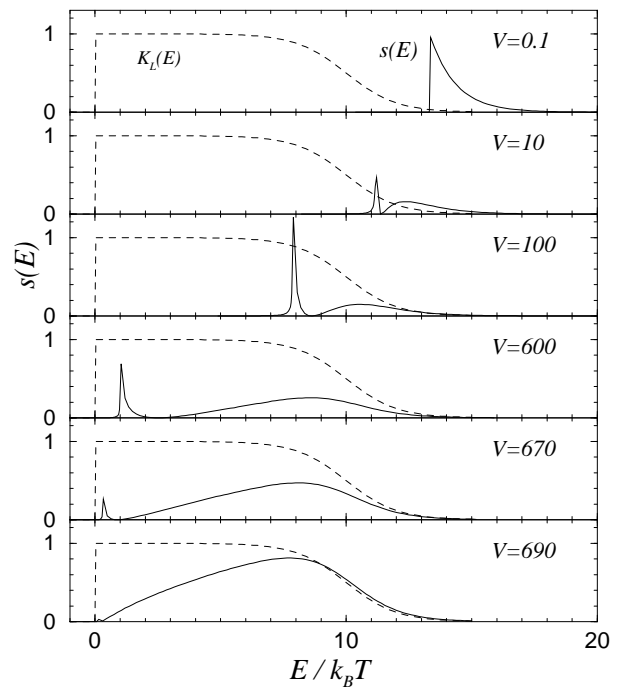}
\caption{Distributions similar to Fig.~\ref{f8} for another ballistic length 
$\ell$=0.5$\mu$m. The notations are the same, except that $K_L(E)$
is not normalized.}
\label{f9} \end{figure}

It is instructive to plot the energy-resolved current-noise power 
$s(E)$ defined by $S_I=\int s(E) dE$.
The derived formula [Eq.~(\ref{SI})] allows us to analyze these distributions 
for different lengths and biases. The results for $\ell$=0.05 and
$\ell$=0.5$\mu$m are shown in Figs.~\ref{f8} and \ref{f9}.
At small biases ($V$=0.1 in the figures), the Coulomb interactions
are ineffective, and the noise is approximately the sum of two equal
contributions from the left and right leads. These contributions are
the Fermi-Dirac profiles filled out above the barrier
(the contributions of the reflected carriers with energies below the barrier
is negligible at these biases).
With increasing the bias, several features appear: (i) the contribution
from the right contact becomes smaller and smaller because of the shift
in energy $-V$, (ii) the Coulomb interactions give rise to a sharp peak
at the barrier energy with a noise suppression at the energies beyond
the peak, and (iii) the carriers below the barrier give appreciable nonzero 
contribution to the noise. 
The peak appears due to the fact that electrons with the energy  
$E=\Phi_L$ virtually stop at the barrier top,
producing an infinitely large perturbation of the current (this singularity 
is integrable, since it is of the logarithmic type \cite{prb00a}).
Another interesting feature is the ``noiseless'' energy $E^*$ lying
above the barrier, in which the noise exhibits a local minimum.
It is better pronounced for a nondegenerate injection (see, for instance, 
Fig.~\ref{f9}) where one can observe the zero-noise point $s(E^*)$=0
for various biases). This point appears approximately at the condition:
$\gamma_L(E^*)$=0.
As long as the barrier vanishes at highest biases $U\to U_{cr}$, the Coulomb 
noise suppression disappears and the energy profile $s(E)$ recovers 
the FD shape.

\subsection{Fixed-current conditions: $\bm{\delta I=0}$}
\label{resultsB}

Thus far we have presented the results obtained under the assumption
that the ballistic sample is connected to zero-impedance external circuit.
In this case the fluctuations of the applied voltage can be neglected.
In experiments, it is the voltage fluctuations which are actually measured
and which eventually are converted to current fluctuations.
By using an infinite-impedance circuit, the current fluctuations are
forced to be zero, and one can analyze the voltage-noise power. 
Both cases are interrelated through Eq.~(\ref{SU1}).
It is of interest however to see the results for the voltage-noise power --
the quantity that can be measured directly. 

\begin{figure}[h]
\epsfxsize=8.0cm
\epsfbox{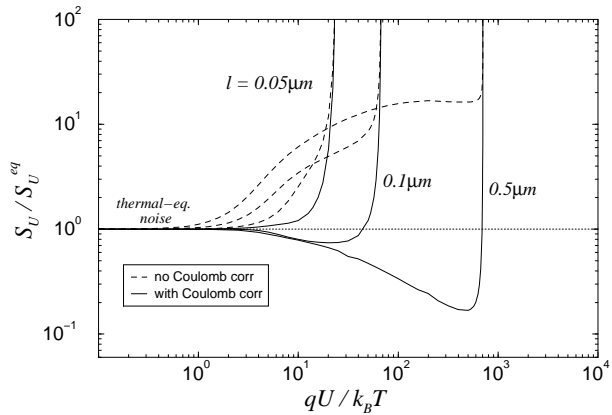}
\caption{Voltage-noise powers $S_U$ vs bias $U$ 
for different ballistic lengths $\ell$ are compared for two cases:
with and without Coulomb correlations taken into account.}
\label{f10} \end{figure}

Figure \ref{f10} shows the results of applying of Eq.~(\ref{SU1}) 
to our set of parameters. The behavior of $S_U$ calculated with and without 
Coulomb correlations is strikingly different. 
We remark the following features:
(i) For the case when the interactions are included, the noise decreases 
with bias instead of increasing [the asymptotic behavior $S_U\sim 1/\sqrt{U}$ 
at high biases (see Sec.~\ref{secSU}) is confirmed].
(ii) For longer samples, the range of the space-charge conduction is wider,
and the suppression of voltage fluctuations is much more pronounced.
(iii) Comparing the asymptotic dependences $S_U\sim 1/\sqrt{U}$ and
$S_I\sim\sqrt{U}$, it is seen that the latter eventually exceeds 
the equilibrium Nyquist noise when the ballistic length is sufficiently long 
[see Fig.~\ref{f4}(c)]. In contrast, $S_U$ falls off the equilibrium value
in a full range of SCL conduction.

Note that the equilibrium fluctuations, described by 
the fluctuation-dissipation theorem, usually increase when an external 
electric field is applied. In our model, we have an interesting example when 
an interacting (via Coulomb forces) electron system is less noisy 
at far-from-equilibrium conditions than in equilibrium.
For noninteracting electrons in quantum conductors, such examples have been
provided by Lesovik and Loosen. \cite{lesovik93} 

We can also mention more familiar examples from semiconductor device 
literature.
For instance, in Schottky-barrier diodes or $p$-$n$ junctions, 
in the range of the exponential $I$-$V$ characteristics,
the current-noise power is given by $S_I=2k_BT G$, which is a half of the 
thermal noise value given by the Nyquist relationship. \cite{ziel86-2} 
However, this is not really a suppression effect, since the current-noise 
power $S_I$ never drops down the equilibrium Nyquist level 
$S_I^{eq}=4k_BT G_{eq}$. 
In this case  $G\gg G_{eq}$ and $S_I>S_I^{eq}$ for any bias \cite{remark4}
(in our model $S_I<S_I^{eq}$ in a wide bias range).
On the other hand, for these junctions under the fixed-current conditions, 
$S_U=2k_BT G^{-1}$, that is again a half of the Nyquist relationship, 
but $S_U<S_U^{eq}$ may now occur. \cite{remark4}
It should be emphasized, however, that the latter noise reduction 
appears for {\em noninteracting} carriers, and it is caused
by the nonlinearity in the current-voltage characteristics which results
in such a behavior that the conductance $G$ grows with bias as fast as the
current-noise power $S_I$ (exponentially). 
As a result, $S_U=S_I/G^2$ is a decreasing function of bias. 
In our model, the noise suppression below the thermal equilibrium value
(negative excess voltage noise) occurs due to {\em Coulomb interactions} 
among carriers.
Without interactions, despite the nonlinear SCL regime, the noise grows
above the Nyquist level, as was shown in Figs.~\ref{f4} and \ref{f10}.

\section{Discussion}

In this paper, we have presented a theory of the electron transport and noise
in a self-consistent potential along a ballistic two-terminal conductor.
Since electrons are fermions and carry charge, they interact among themselves 
by both Fermi statistical correlations and long-range Coulomb correlations.
The interplay of these two mechanisms determine the noise properties 
of a ballistic conductor, -- the subject we have addressed in the paper.

The long-range Coulomb correlations appear due to the self-consistent
coupling between the electric potential and the occupation numbers.
This coupling is essential to adequately describe the noise phenomena.
To develop a better understanding of the Coulomb-correlation effect,
we rewrite Eq.~(\ref{mcur1})---a standard equation for the mean 
current in a two-terminal conductor---in which we explicitly introduce 
the dependence of the transmission probabilities on the self-consistent 
potential (in this case on the barrier height $\Phi_b$):
\begin{equation}  \label{mcur2}
I = \int d\varepsilon\, [{\cal T}_L(\varepsilon,\Phi_b)\, I_L(\varepsilon)
- {\cal T}_R(\varepsilon,\Phi_b)\, I_R(\varepsilon)].
\end{equation}
Here, $I_{L,R}(\varepsilon)=(qA/2\pi\hbar)\int 
f_{L,R}(\varepsilon,\varepsilon_{\perp}) \nu_{\perp} d\varepsilon_{\perp}$
are the currents corresponding to the longitudinal energy $\varepsilon$.
In the semiclassical limit, by neglecting the quantum-mechanical reflection 
of electrons with energies $\varepsilon>\Phi_b$ and tunneling through 
the barrier, the transmission probabilities are the Heaviside step functions:
${\cal T}_L(\varepsilon,\Phi_b)=\theta(\varepsilon-\Phi_b)$
and 
${\cal T}_R(\varepsilon,\Phi_b)=\theta(\varepsilon-\Phi_b-qU)$.

The current fluctuation is found by perturbing Eq.~(\ref{mcur2}); we obtain
\begin{align}  \label{dcur}
&\delta I = 
\int [{\cal T}_L\ \delta I_L- {\cal T}_R\ \delta I_R]\, d\varepsilon 
\nonumber\\ &
+\int \left[ (- \delta\Phi_b) \, 
\frac{\partial {\cal T}_L}{\partial \varepsilon}\ I_L\,
+ (-\delta\Phi_b-q\delta U) \, 
\frac{\partial {\cal T}_R}{\partial \varepsilon}\ I_R \, 
\right] d\varepsilon \nonumber\\
&\equiv \delta I_{inj} + \delta I_{ind}, 
\end{align}
The first integral in Eq.~(\ref{dcur}) is the injected current fluctuation
$\delta I_{inj}$. Its origin is the thermal fluctuation of the occupation
numbers in the leads ballistically injected into the conductor.
In fact this term corresponds to Eq.~(\ref{Stwo}) -- the standard formula
used to calculate the current noise in mesoscopic conductors.
[Since for our case all ${\cal T}_n=0$ or 1, 
the term $\propto {\cal T}_n(1-{\cal T}_n)$ is absent.]
The second integral in Eq.~(\ref{dcur}) is the induced current fluctuation
$\delta I_{ind}$ caused by the fluctuation of the potential. 
It is precisely the term appeared due to Coulomb correlations and 
ignored in Eq.~(\ref{Stwo}).
For our case, the derivatives are found as:
$(\partial {\cal T}_L/\partial \varepsilon)=\delta(\varepsilon-\Phi_b)$,
$(\partial {\cal T}_R/\partial \varepsilon)=\delta(\varepsilon-\Phi_b-qU)$,
leading to
\begin{equation}  \label{dcurind}
\delta I_{ind} =  - \delta\Phi_b \,I_L(\Phi_b) 
- (\delta\Phi_b-q\delta U)\, I_R(\Phi_b+qU).
\end{equation}
Thus Eqs.~(\ref{dcur}) and (\ref{dcurind}) lead to Eq.~(\ref{dI}) for the 
current fluctuation derived more rigorously earlier 
from the transport equation.

We would like to highlight that the induced current fluctuations 
$\delta I_{ind}$ should appear not only in the case of completely open/closed
channels (${\cal T}_n=0;1$), but also under the conditions of 
the partitioning shot noise, for which there exist channels with 
$0<{\cal T}_n<1$.
It is clear that $\delta I_{ind}$ should depend in general on the derivatives
of the transmission probabilities $(\partial {\cal T}_n/\partial \varepsilon)$
and the fluctuations of the self-consistent potential $\delta\Phi_x$. 
The main problem is then to find the fluctuations $\delta\Phi_x$
through the noise sources.
For the particular case of a multimode ballistic two-terminal conductor,
we have found an exact analytical result for $\delta\Phi_x$.
For the case of partitioning shot noise in which Eq.~(\ref{Stwo}) holds, 
work is in progress. 

The validity of our theory can be tested experimentally in currently 
accessible semiconductor structures. The required conditions are similar 
\cite{remark5} to those for the transport in vacuum tubes:
(i) the ballistic electron transmission
between the terminals, and (ii) the limitation of current by the space charge.
The SCL transport regime, as applied for ballistic electrons in solids
(mostly in $n^+$-$n$-$n^+$ or $n$-$i$-$n$ semiconductor structures),
was discussed a long time ago (see, e.g., theory \cite{SCL-ball-theory} 
and experiments \cite{SCL-ball-exp}). Unfortunately, the data on 
noise measurements in these structures are scarce. \cite{schmidt83}
Due to a great progress in noise measurements in quantum ballistic conductors 
during the last ten years \cite{li90,reznikov95,kumar96,brom99} 
(also see Ref.~\onlinecite{blanter00}),
we believe it would be now possible to measure the noise suppression
effects in SCL ballistic conductors.
Although the theoretical results presented in this paper are strictly valid 
for thick multichannel conductors (3D electron gas), the Coulomb suppression 
of noise should also be pronounced \cite{land-martin9192} in conductors 
with a small number of channels (2D or 1D) in which electrons are more 
confined in space, for instance, in quantum wires under the high-bias 
nonlinear transport regime, \cite{novoselov00} or
in carbon nanotubes under the SCL conduction. \cite{nanotubes}

Additionally, we would like to emphasize the importance of the effect of
Coulomb interactions. They not only lead to the noise reduction,
but can also be used as a tool to probe the energy profile of the injected
carriers and other electronic properties. \cite{pe02}

\begin{acknowledgements}
This work was partially supported by the Ministerio de Ciencia y 
Tecnolog\'{\i}a of Spain through the ``Ram\'on y Cajal'' program.
\end{acknowledgements}

\appendix
\section{Derivation of $\bm{dU_b/dU}$}
\label{app-g}

Differentiating Eq.~(\ref{Vb}) gives
\begin{multline} \label{diff1}
\frac{d\Phi_L}{dU} \frac{1}{\sqrt{h_-(\Phi_L)}}
-\frac{1}{2} \int_0^{\Phi_L} \frac{1}{h_-^{3/2}} \frac{dh_-}{dU}\,d\Phi \\
+\frac{d\Phi_R}{dU} \frac{1}{\sqrt{h_+(\Phi_R)}} 
-\frac{1}{2} \int_0^{\Phi_R} \frac{1}{h_+^{3/2}} \frac{dh_+}{dU}\,d\Phi=0.
\end{multline}
By using $d\Phi_R/dU=q+d\Phi_L/dU$ and finding $dh/dU$ from Eq.~(\ref{h}),
we obtain
\begin{equation} \label{dh}
\frac{dh}{dU} = - (H_L + H_R)\ \frac{d\Phi_L}{dU} - q H_R,
\end{equation}
where 
\begin{eqnarray} \label{H}
H_L(\Phi) &=& \frac{m}{2\pi\hbar}  \left\{ 
\int_0^{\infty} {\cal D}_L\ [v(E+\Phi)-v(E)]\ dE \right.
\nonumber\\ &&
\left. + 2 \theta(-\chi) \int_{-\Phi}^0 {\cal D}_L\ v(E+\Phi)\ 
dE \right\}, \\
H_R(\Phi) &=& \frac{m}{2\pi\hbar}  \left\{ 
\int_0^{\infty} {\cal D}_R\
[v(E+\Phi)-v(E)]\ dE \right.  \nonumber\\ &&
\left. + 2 \theta(\chi)  \int_{-\Phi}^0 
{\cal D}_R\ v(E+\Phi)\ dE \right\}, \label{HR} \\
{\cal D}_k(E) &=&  \left. -\frac{\partial n_k}{\partial E} 
\right|_{E+\Phi_k}, \quad k=L,R. \label{calD}
\end{eqnarray}
Substituting Eqs.~(\ref{dh})--(\ref{HR}) into (\ref{diff1}), we finally obtain
\begin{equation} \label{dudum}
\frac{dU_b}{dU} = \frac{1}{q} \frac{d\Phi_L}{dU} 
= -\frac{\Delta_R}{\Delta_L + \Delta_R}.
\end{equation}
Here we have defined
\begin{eqnarray*} \label{deltaR}
\Delta_L &=& \frac{2}{\sqrt{h_-(\Phi_L)}}
+\int_0^{\Phi_L} \frac{H_L^-}{h_-^{3/2}} d\Phi 
+\int_0^{\Phi_R} \frac{H_L^+}{h_+^{3/2}} d\Phi, \\
\Delta_R &=& \frac{2}{\sqrt{h_+(\Phi_R)}}
+\int_0^{\Phi_L} \frac{H_R^-}{h_-^{3/2}} d\Phi 
+\int_0^{\Phi_R} \frac{H_R^+}{h_+^{3/2}} d\Phi,
\end{eqnarray*}
where we have denoted $H_k^-\equiv H_k(\chi<0)$ and $H_k^+\equiv H_k(\chi>0)$, 
$k=L,R$.
The quantities $\Delta_L$ and $\Delta_R$, as well as their sum
\begin{equation} \label{delta}
\Delta = \Delta_L + \Delta_R,
\end{equation}
are used in this paper to calculate the differential conductance
[Eq.~(\ref{g1})] and the noise suppression [Eq.~(\ref{Cdelta})].
The physical meaning of $\Delta$ becomes clear from the relation
\begin{equation}
\Delta = 2q \sqrt{\frac{2}{\kappa}}
\left( \frac{\partial\ell}{\partial\Phi_L} \right)_{U=\text{const}},
\end{equation}
i.e., it relates the increment of the barrier height with the increase
of the length of the sample under fixed bias.
$\Delta_{L,R}$ are the corresponding contributions to that increment
from the left-lead and right-lead electrons.

\section{Derivation of the self-consistent potential fluctuations}
\label{app-pot}

Integrating the fluctuation of the occupation factor $\delta n(E)$ 
over the longitudinal states, one obtains the electron-density fluctuation 
as a sum of two contributions,
$\delta N=\delta N^{inj}+\delta N^{ind}$,
where the injected part is
\begin{widetext}
\begin{multline} \label{dninj}
\delta N^{inj}(\Phi) =
\int_0^{\infty} [\delta n_L(E+\Phi_L)+
\delta n_R(E+\Phi_R)] \nu(E+\Phi) dE \\
+ 2 \int_{-\Phi}^0 [\theta(-\chi)\delta n_L(E+\Phi_L)
+\theta(\chi) \delta n_R(E+\Phi_R)] \nu(E+\Phi) dE,
\end{multline}
\end{widetext}
and the induced part is
\begin{eqnarray} \label{dnind}
\delta N^{ind}(\Phi) =
\frac{dN}{d\Phi}\delta\Phi_x 
- \frac{dH_L}{d\Phi} \delta\Phi_L 
- \frac{dH_R}{d\Phi} \delta\Phi_R.
\end{eqnarray}
Equations (\ref{dninj}) and (\ref{dnind}) should now be substituted into 
the Poisson equation for $\delta\Phi_x$ to find the self-consistent 
fluctuation of the potential profile:
\begin{eqnarray} \label{dpois}
\frac{d^2}{dx^2} \delta\Phi_x = \frac{q^2}{\kappa} \,
(\delta N^{inj} + \delta N^{ind}).
\end{eqnarray}
We get
\begin{eqnarray} \label{nhom}
\hat{L}\delta\Phi_x&\equiv& 
\left[\frac{d^2}{dx^2} - \frac{q^2}{\kappa} \frac{dN}{d\Phi} \right] 
\delta\Phi_x = \delta s_x,
\end{eqnarray}
where $\delta s_x = (q^2/\kappa)
[\delta N^{inj} - (dH_L/d\Phi)\delta\Phi_L - (dH_R/d\Phi)\delta\Phi_R]$.
The boundary conditions for this equation
$\delta\Phi_x(0)=\delta\Phi_L$, $\delta\Phi_x(\ell)=\delta\Phi_R$,
$\delta\Phi_x(x_b)=0$.

The second-order differential equation (\ref{nhom}) with spatially dependent 
coefficients can be solved explicitly for $\delta\Phi_x$. \cite{prb00a}
Here we need just the boundary values $\delta\Phi_L$ and $\delta\Phi_R$ 
(the relation between them), which has entered explicitly 
into the nonhomogeneous part and can be obtained 
by applying the Green's identity for the operator $\hat{L}$,
\begin{eqnarray} \label{green}
\int_a^b [u(x) \hat{L}\delta\Phi_x &-& \delta\Phi_x\hat{L}u(x)] dx
\nonumber\\
&=& \left. \left( u(x)\frac{d\delta\Phi}{dx} - \delta\Phi_x \frac{du}{dx}
\right)\right|_a^b,
\end{eqnarray}
where $[a;b]=[0;x_b]$ for $\chi<0$ and $[a;b]=[x_b;\ell]$ for $\chi>0$.
It is convenient to chose the function $u(x)$ as a solution of the 
homogeneous equation $\hat{L}u(x)$=0 satisfying the boundary conditions
$u(0)$=0 and $u(l)$=0. This gives
\begin{eqnarray*} \label{green2}
\int_0^{x_b} u \, \delta s_x \, dx + \int_{x_b}^{\ell} u \, \delta s_x \, dx
= \frac{\delta\Phi_L}{{\cal E}_L} - \frac{\delta\Phi_R}{{\cal E}_R},
\end{eqnarray*}
where ${\cal E}_L$ and ${\cal E}_R$ are the electric fields at $x$=0 and
$x$=$\ell$, respectively.
Changing the variable of integration $dx$=$-d\Phi/(q{\cal E})$ , one gets
\begin{eqnarray} \label{green3}
\int_0^{\Phi_L} \frac{u}{{\cal E}}\, \delta s_x \, d\Phi -
\int_0^{\Phi_R} \frac{u}{{\cal E}}\, \delta s_x \, d\Phi 
= \frac{\delta\Phi_L}{{\cal E}_L} - \frac{\delta\Phi_R}{{\cal E}_R}.
\end{eqnarray}
It is convenient to represent the fluctuation $\delta s_x$
as a derivative $\delta s_x = (\kappa/q^2)\,(d\delta h/d\Phi)$. 
By using this notation, the integrals in Eq.~(\ref{green3}) can be reduced to
\cite{prb00b}
\begin{eqnarray}\label{green4}
\int_0^{\Phi_k} \frac{u}{{\cal E}}\, \frac{d\,\delta h}{d\Phi}\, d\Phi =
\frac{1}{q} \int_0^{\Phi_k} \frac{\delta h}{{\cal E}^3} \, d\Phi, \quad k=L,R
\end{eqnarray}
whereas $\delta h$ is obtained by integration of $\delta s_x$:
\begin{eqnarray} \label{dY}
&&\delta h = \delta h^{inj} - H_L\delta\Phi_L - H_R\,\delta\Phi_R, \\
&&\delta h^{inj}(\Phi) = \int_0^{\Phi}\delta N^{inj} d\tilde{\Phi}.
 \label{dYinj}
\end{eqnarray}
Now substituting Eqs.~(\ref{green4}) and (\ref{dY}) into Eq.~(\ref{green3}),
and by using Eq.~(\ref{e}), we obtain
\begin{eqnarray} \label{rel1}
\Delta \delta\Phi_L + \Delta_R q\delta U = 
\int_0^{\Phi_L} \frac{\delta h_-^{inj}}{h_-^{3/2}} \, d\Phi
+\int_0^{\Phi_R} \frac{\delta h_+^{inj}}{h_+^{3/2}} \, d\Phi, \nonumber\\
\end{eqnarray}
where $\Delta$ and $\Delta_R$ were denoted in Appendix \ref{app-g}. Combining 
Eqs.~(\ref{dI}) and (\ref{rel1}) and excluding $\delta\Phi_L$, we obtain
\begin{multline} \label{rel2}
\delta I - G \delta U =
 \int_0^{\infty} [\delta I_L(E+\Phi_L)-\delta I_R(E+\Phi_R)]
\, d E \nonumber\\
- \frac{I_L(\Phi_L)-I_R(\Phi_R)}{\Delta} \left(
\int_0^{\Phi_L} \frac{\delta h_-^{inj}}{h_-^{3/2}} \, d\Phi
+\int_0^{\Phi_R} \frac{\delta h_+^{inj}}{h_+^{3/2}}\, d\Phi \right),
\end{multline}
that leads to Eq.~(\ref{rel3}).

\section{Nyquist theorem and the boundary conditions for fluctuations}
\label{app-N}

Consider the situation when the potentials at the leads are held equal 
($U$=0, $\delta U$=0) by means of a zero-impedance external circuit. 
(A similar consideration can be carried out for the infinite-impedance
circuit.)
Additionally we assume that the contacts are identical:
$I_L(E)=I_R(E)$, and $K_L(E)=K_R(E)$, $\forall E$.
Thus from Eq.~(\ref{Cdelta}) we have $C_{\Delta}$=0, $\gamma_L(E)=\theta(E)$, 
and $\gamma_R(E)=-\theta(E)$,
which means that Coulomb correlations do not affect noise at zero bias.
Therefore, from Eqs.~(\ref{rel3}) and (\ref{kuncor}) one obtains 
the current-noise power
\begin{equation} \label{SIeq}
S_I^{eq} = 2\int_0^{\infty} K_L(E+\Phi_L^0)\,dE,
\end{equation}
where $\Phi_L^0$ is the equilibrium barrier height
(the noise depends on the steady-state self-consistent field).
For the equilibrium conductance we find
\begin{eqnarray} \label{g0}
G_{eq} = \left. \frac{dI}{dU}\right|_{U=0} &=& -q \int_0^{\infty} \left.
\frac{\partial I_L}{\partial E}\right|_{E+\Phi_L^0} dE.
\end{eqnarray}
By using the Nyquist theorem $S_I^{eq} = 4 k_B T G_{eq}$, we obtain
\begin{eqnarray*} 
\int_0^{\infty} K_L(E+\Phi_L^0)\,dE = 2q k_B T 
\int_0^{\infty} \left.\left(-\frac{\partial I_L}{\partial E}
\right)\right|_{E+\Phi_L^0} dE.
\end{eqnarray*}
Since this integral relation should be valid for different lengths $\ell$ 
of the ballistic conductor (different $\Phi_L^0$),
it should also be valid for the integrands,
\begin{equation} \label{KN}
K_L(E) = 2 q k_B T  
\left(-\frac{\partial I_L}{\partial E}\right),
\end{equation}
that leads to Eq.~(\ref{K}).
Thus, just from the Nyquist theorem we have a useful relation 
for the energy-resolved currents (occupation factors) at the leads.
It relates the energy profiles of the fluctuations and the mean values.
In the simplest case of the Poissonian injection, for instance, 
the correlation function is proportional to the mean \cite{kogan}
\begin{equation}
K_L^{\text{Pois}}(E)=2qI_L(E).
\end{equation}
{}From this result it follows that 
$I_L(E)=-k_BT (\partial I_L/\partial E)$,
and one obtains the Boltzmann distribution
\begin{equation} \label{IL_Pois}
I_L^{\text{Pois}}(E) = C \exp(-E/k_BT),
\end{equation}
where the integration constant $C$ is determined by the normalization 
condition.


\begin{thebibliography}{}

\bibitem{dejong97}
M.J.M.~de Jong and C.W.J.~Beenakker,
in {\it Mesoscopic Electron Transport},
edited by L.P.~Kowenhoven, G.~Sch\"on, and L.L.~Sohn
(Kluwer, Dordrecht, 1997), p.~225.

\bibitem{landauer98}
R.~Landauer, Nature (London) {\bf 392}, 658 (1998).

\bibitem{blanter00}
Ya.M.~Blanter and M.~B\"uttiker, Phys. Rep. {\bf 336}, 1 (2000).

\bibitem{lesovik89} 
G.B.~Lesovik, Pis'ma Zh. \'Eksp. Teor. Fiz. {\bf 49}, 513 (1989) 
[JETP Lett. {\bf 49}, 592 (1989)]. 

\bibitem{buttiker9092} 
M.~B\"uttiker, Phys. Rev. Lett. {\bf 65}, 2901 (1990);
Phys. Rev. B {\bf 46}, 12485 (1992). 

\bibitem{land-martin9192} 
R.~Landauer and Th.~Martin, Physica B {\bf 175}, 167 (1991);
Th.~Martin and R.~Landauer, Phys. Rev. B {\bf 45}, 1742 (1992).

\bibitem{li90} 
Y.P.~Li, D.C.~Tsui, J.J.~Heremans, J.A.~Simmons,
and G.W.~Weimann, Appl. Phys. Lett. {\bf 57}, 774 (1990). 

\bibitem{reznikov95}
M.~Reznikov, M.~Heiblum, H.~Shtrikman, and D.~Mahalu,
Phys. Rev. Lett. {\bf 75}, 3340 (1995).

\bibitem{kumar96}
A.~Kumar, L.~Saminadayar, D.C.~Glattli, Y.~Jin, and B.~Etienne,
Phys. Rev. Lett. {\bf 76}, 2778 (1996).

\bibitem{brom99} 
H.E.~van den Brom and J.M.~van Ruitenbeek, Phys. Rev.
Lett. {\bf 82}, 1526 (1999).

\bibitem{kogan}
Sh.~Kogan, {\it Electronic Noise and Fluctuations in Solids}
(Cambridge University Press, Cambridge, 1996), Chap.~5.

\bibitem{remark1}
In Eq.~(\ref{Stwo1}), $S_I$ depends on the bias through the parameter $n^*$, 
i.e., the number of open channels which may vary with bias. 
The same is true for the current. In the absence of partitioning, 
the current varies with bias due to the space-charge impediment.

\bibitem{landauer89}
R.~Landauer, Physica D {\bf 38}, 226 (1989).

\bibitem{sablikov98} 
See, for instance, the recent study by
V.A.~Sablikov and B.S.~Shchamkhalova, Phys. Rev. B {\bf 58}, 13847 (1998).

\bibitem{sablikov00} 
V.A.~Sablikov, S.V.~Polyakov, and M.~B\"uttiker, 
Phys. Rev. B {\bf 61}, 13763 (2000).

\bibitem{buttiker02} 
M.~B\"uttiker, Pramana-J. Phys. {\bf 58}, 241 (2002).

\bibitem{wei99} 
Y.~Wei, B.~Wang, J.~Wang, and H.~Guo, Phys. Rev. B {\bf 60}, 16900 (1999).

\bibitem{prb00a}
O.M.~Bulashenko, J.M.~Rub\'{\i}, and V.A.~Kochelap,
Phys. Rev. B {\bf 61}, 5511 (2000).

\bibitem{landauer9396}
R.~Landauer, Phys. Rev. B {\bf 47}, 16427 (1993);
Physica B {\bf 227}, 156 (1996).

\bibitem{buttiker96}
M.~B\"uttiker, J. Math. Phys. {\bf 37}, 4793 (1996).

\bibitem{lesovik93} 
G.B.~Lesovik and R.~Loosen, Z. Phys. B {\bf 91}, 531 (1993).

\bibitem{scherbakov98} 
A.G.~Scherbakov, E.N.~Bogachek, and U.~Landman, 
Phys. Rev. B {\bf 57}, 6654 (1998).

\bibitem{levi85}
A.F.J.~Levi, J.R.~Hayes, P.M.~Platzman, and W.~Wiegmann,
Phys. Rev. Lett. {\bf 55}, 2071 (1985).

\bibitem{heiblum85}
M.~Heiblum, M.I.~Nathan, D.C.~Thomas, and C.M.~Knoedler,
Phys. Rev. Lett. {\bf 55}, 2200 (1985).

\bibitem{kast01}
M.~Kast, C.~Pacher, G.~Strasser, and E.~Gornik, 
Appl. Phys. Lett. {\bf 78}, 3639 (2001).

\bibitem{strahberger01}
C.~Strahberger, J.~Smoliner, R.~Heer, and G.~Strasser,
Phys. Rev. B {\bf 63}, 205306 (2001).

\bibitem{prb00b}
O.M.~Bulashenko, J.M.~Rub\'{\i}, and V.A.~Kochelap,
Phys. Rev. B {\bf 62}, 8184 (2000).

\bibitem{remark2}
Equation (\ref{Vb}) 
is valid for the biases $U<U_{cr}$ when the barrier exists, 
where $U_{cr}$ is the bias for which $d\Phi/dx$=0 at $x$=0.
The case $U>U_{cr}$ is not interesting, since the transport is no longer 
space-charge limited, and the noise is no longer suppressed by Coulomb 
correlations.

\bibitem{prb01}
O.M.~Bulashenko and J.M.~Rub\'{\i}, 
Phys. Rev. B {\bf 64}, 045307 (2001).


\bibitem{ziel86}
A.~van der Ziel, {\it Noise in Solid State Devices and Circuits},
(Wiley, New York, 1986), Chap.~3.1.

\bibitem{shiktorov96}
P.~Shiktorov, V.~Gruzinskis, E.~Starikov, L.~Reggiani, and L.~Varani,
Phys. Rev. B {\bf 54}, 8821 (1996).

\bibitem{apl99}
O.M.~Bulashenko, J.M.~Rub\'{\i}, and V.A.~Kochelap,
Appl. Phys. Lett. {\bf 75}, 2614 (1999).

\bibitem{remark3}
In $n$-$i$-$n$ structures, the current is mostly carried by electrons 
injected from the contacts ($n$ regions). 
The contribution of intrinsic carriers of the sample ($i$ region)
is negligible.
Similar calculations can be carried out for $n^+$-$n$-$n^+$,
$n^+$-$p$-$n^+$ diodes, heterodiodes (Ref.~\onlinecite{apl99}) and others.

\bibitem{ziel86-2}
A.~van der Ziel, {\it Noise in Solid State Devices and Circuits},
(Wiley, New York, 1986), Chap.~6.1.

\bibitem{remark4}
In Schottky-barrier diodes and $p$-$n$ junctions, the mean current is a sum
of two opposite current flows crossing the barrier: $I=I_L-I_R$, with
$I_L=I_0\exp(qU/k_BT)$ and $I_R=I_0$ (see Ref.~\onlinecite{ziel86-2}). Hence 
the conductance is $G=q(I+I_0)/k_BT$, assuming $I_0$ is a slow function of $U$.
Neglecting correlations among carriers, the current noise power is a sum
of two Poissonian noise terms: 
$S_I=2q(I_L+I_R)=2q(I+2I_0)=2k_BT\,G[(I+2I_0)/(I+I_0)]$.
The nonequilibrium noise at high currents $I\gg I_0$ is $S_I=2k_BT\,G$, 
which is a half of the Nyquist thermal noise of a resistor having the
identical (nonequilibrium) conductance $G$. However, by rewriting 
$S_I=S_I^{eq}[1+I/(2I_0)]$, it is easy to see that $S_I>S_I^{eq}$ for any $I$.
On the other hand, the voltage-noise power 
$S_U=S_I/G^2=S_U^{eq}[(1+I/2I_0)/(1+I/I_0)^2]$ is lower than the equilibrium 
Nyquist value $S_U^{eq}$, since it decreases with current as 
$S_U\approx S_U^{eq} (I_0/2I)$.

\bibitem{remark5}
The distinction between the electron transport in ballistic solid-state and 
vacuum diodes was drawn in Refs.\ \onlinecite{prb00a,apl99}. 

\bibitem{SCL-ball-theory}
M.S.~Shur and L.F.~Eastman, 
IEEE Trans. Electron. Devices {\bf 26}, 1677 (1979);
A.J.~Holden and B.T.~Debney, Electron. Lett. {\bf 18}, 558 (1982);
H.U.~Baranger and J.W.~Wilkins, Phys. Rev. B {\bf 30}, 7349 (1984).

\bibitem{SCL-ball-exp}
L.F.~Eastman, R.~Stall, D.~Woodard, N.~Dandekar, C.E.C.~Wood, M.S.~Shur,
and K.~Board, Electron. Lett. {\bf 16}, 524 (1980);
M.A.~Hollis,  L.F.~Eastman, and C.E.C.~Wood, 
Electron. Lett. {\bf 18}, 524 (1982); more recently,
E.~Hern\'andez, Cryst. Res. Technol. {\bf 33}, 285 (1998).

\bibitem{schmidt83}
R.R.~Schmidt, G.~Bosman, C.M.~Van Vliet, L.F.~Eastman, and M.~Hollis,
Solid-State Electron. {\bf 26}, 437 (1983).

\bibitem{novoselov00}
K.S.~Novoselov, Yu.V.~Dubrovskii, V.A.~Sablikov, D.Yu.~Ivanov,
E.E.~Vdovin, Yu.N.~Khanin, V.A.~Tulin, D.~Esteve, and S.~Beaumont,
Europhys. Lett. {\bf 52}, 660 (2000).

\bibitem{nanotubes}
Z.K.~Tang, H.D.~Sun, J.~Wang, J.~Chen, and G.~Li,
Appl. Phys. Lett. {\bf 73}, 2287 (1998);
Ph.~Avouris, R.~Martel, H.~Ikeda, M.~Hersam, H.R.~Shea, and A.~Rochefort,
in {\it Science and Application of Nanotubes}, edited by D.~Tom\'anek
and R.J.~Enbody (Kluwer, New York, 2000) p.~223.





\bibitem{pe02}
O.M.~Bulashenko and J.M.~Rub\'{\i}, Physica E {\bf 12}, 857 (2002).

\end{thebibliography}
\end{document}